\documentclass[useAMS,usenatbib,usegraphicx]{mn2e}

\title[Hydrodynamics and X-Ray Emission from SN 1996cr]{Bursting SN
  1996cr's Bubble: Hydrodynamic and X-ray Modeling of its
  Circumstellar Medium} \author[Dwarkadas et
  al.]{V. V. Dwarkadas,$^{1}$\thanks{E-mail:
    vikram@oddjob.uchicago.edu}
  D. Dewey,$^{2}$\thanks{dd@space.mit.edu}F. Bauer$^{3,4,5}$\thanks{fbauer@spacescience.org}\\ $^{1}$Department
  of Astronomy and Astrophysics, U Chicago, 5640 S Ellis Ave, Chicago,
  IL 60637\\ $^{2}$MIT Kavli Institute, Massachusetts Institute of
  Technology, Cambridge, MA 02139\\ $^{3}$Space Science Institute,
  4750 Walnut Street, Suite 205, Boulder, Colorado
  80301\\$^{4}$Pontificia Universidad Cat\'{o}lica de Chile,
  Departamento de Astronom\'{\i}a y Astrof\'{\i}sica, Casilla 306,
  Santiago 22, Chile\\ $^{5}$Columbia Astrophysics Laboratory, 550
  W. 120th St., Columbia University, New York, NY 10027}

\begin{document}
\newcommand{\vper}{\mbox{${v_{\perp}}$}}
\newcommand{\vpar}{\mbox{${v_{\parallel}}$}}
\newcommand{\uper}{\mbox{${u_{\perp}}$}}
\newcommand{\vperout}{\mbox{${{v_{\perp}}_{o}}$}}
\newcommand{\uperout}{\mbox{${{u_{\perp}}_{o}}$}}
\newcommand{\vperin}{\mbox{${{v_{\perp}}_{i}}$}}
\newcommand{\uperin}{\mbox{${{u_{\perp}}_{i}}$}}
\newcommand{\upar}{\mbox{${u_{\parallel}}$}}
\newcommand{\uparout}{\mbox{${{u_{\parallel}}_{o}}$}}
\newcommand{\vparout}{\mbox{${{v_{\parallel}}_{o}}$}}
\newcommand{\uparin}{\mbox{${{u_{\parallel}}_{i}}$}}
\newcommand{\vparin}{\mbox{${{v_{\parallel}}_{i}}$}}
\newcommand{\dout}{\mbox{${\rho}_{o}$}}
\newcommand{\din}{\mbox{${\rho}_{i}$}}
\newcommand{\da}{\mbox{${\rho}_{1}$}}
\newcommand{\mfast}{\mbox{$\dot{M}_{f}$}}
\newcommand{\mslow}{\mbox{$\dot{M}_{a}$}}
\newcommand{\beqn}{\begin{eqnarray}}
\newcommand{\eeqn}{\end{eqnarray}}
\newcommand{\be}{\begin{equation}}
\newcommand{\ee}{\end{equation}}
\newcommand{\noi}{\noindent}
\newcommand{\ftheta}{\mbox{$f(\theta)$}}
\newcommand{\gtheta}{\mbox{$g(\theta)$}}
\newcommand{\ltheta}{\mbox{$L(\theta)$}}
\newcommand{\stheta}{\mbox{$S(\theta)$}}
\newcommand{\utheta}{\mbox{$U(\theta)$}}
\newcommand{\xitheta}{\mbox{$\xi(\theta)$}}
\newcommand{\vs}{\mbox{${v_{s}}$}}
\newcommand{\ro}{\mbox{${R_{0}}$}}
\newcommand{\pa}{\mbox{${P_{1}}$}}
\newcommand{\va}{\mbox{${v_{a}}$}}
\newcommand{\vo}{\mbox{${v_{o}}$}}
\newcommand{\vp}{\mbox{${v_{p}}$}}
\newcommand{\vw}{\mbox{${v_{w}}$}}
\newcommand{\vf}{\mbox{${v_{f}}$}}
\newcommand{\lprime}{\mbox{${L^{\prime}}$}}
\newcommand{\uprime}{\mbox{${U^{\prime}}$}}
\newcommand{\sprime}{\mbox{${S^{\prime}}$}}
\newcommand{\xiprime}{\mbox{${{\xi}^{\prime}}$}}
\newcommand{\mdot}{\mbox{$\dot{M}$}}
\newcommand{\msun}{\mbox{$M_{\odot}$}}
\newcommand{\yr}{\mbox{${\rm yr}^{-1}$}}
\newcommand{\kms}{\mbox{${\rm km} \;{\rm s}^{-1}$}}
\newcommand{\lambdav}{\mbox{${\lambda}_{v}$}}
\newcommand{\lequ}{\mbox{${L_{eq}}$}}
\newcommand{\eqpratio}{\mbox{${R_{eq}/R_{p}}$}}
\newcommand{\ra}{\mbox{${r_{o}}$}}
\newcommand{\bfig}{\begin{figure}[h]}
\newcommand{\efig}{\end{figure}}
\newcommand{\tone}{\mbox{${t_{1}}$}}
\newcommand{\done}{\mbox{${{\rho}_{1}}$}}
\newcommand{\dsn}{\mbox{${\rho}_{SN}$}}
\newcommand{\dzero}{\mbox{${\rho}_{0}$}}
\newcommand{\ve}{\mbox{${v}_{e}$}}
\newcommand{\vej}{\mbox{${v}_{ej}$}}
\newcommand{\Mch}{\mbox{${M}_{ch}$}}
\newcommand{\mej}{\mbox{${M}_{e}$}}
\newcommand{\Mst}{\mbox{${M}_{ST}$}}
\newcommand{\dam}{\mbox{${\rho}_{am}$}}
\newcommand{\Rst}{\mbox{${R}_{ST}$}}
\newcommand{\Vst}{\mbox{${V}_{ST}$}}
\newcommand{\Tst}{\mbox{${T}_{ST}$}}
\newcommand{\no}{\mbox{${n}_{0}$}}
\newcommand{\Efif}{\mbox{${E}_{51}$}}
\newcommand{\rsh}{\mbox{${R}_{sh}$}}
\newcommand{\msh}{\mbox{${M}_{sh}$}}
\newcommand{\vsh}{\mbox{${V}_{sh}$}}
\newcommand{\vrev}{\mbox{${v}_{rev}$}}
\newcommand{\rpr}{\mbox{${R}^{\prime}$}}
\newcommand{\mpr}{\mbox{${M}^{\prime}$}}
\newcommand{\vpr}{\mbox{${V}^{\prime}$}}
\newcommand{\tpr}{\mbox{${t}^{\prime}$}}
\newcommand{\cone}{\mbox{${c}_{1}$}}
\newcommand{\ctwo}{\mbox{${c}_{2}$}}
\newcommand{\cthree}{\mbox{${c}_{3}$}}
\newcommand{\cfour}{\mbox{${c}_{4}$}}
\newcommand{\Te}{\mbox{${T}_{e}$}}
\newcommand{\Ti}{\mbox{${T}_{i}$}}
\newcommand{\Ha}{\mbox{${H}_{\alpha}$}}
\newcommand{\Rprime}{\mbox{${R}^{\prime}$}}
\newcommand{\Vprime}{\mbox{${V}^{\prime}$}}
\newcommand{\Tprime}{\mbox{${T}^{\prime}$}}
\newcommand{\Mprime}{\mbox{${M}^{\prime}$}}
\newcommand{\rprime}{\mbox{${r}^{\prime}$}}
\newcommand{\rfprime}{\mbox{${r}_f^{\prime}$}}
\newcommand{\vprime}{\mbox{${v}^{\prime}$}}
\newcommand{\tprime}{\mbox{${t}^{\prime}$}}
\newcommand{\mprime}{\mbox{${m}^{\prime}$}}
\newcommand{\Me}{\mbox{${M}_{e}$}}
\newcommand{\nh}{\mbox{${n}_{H}$}}
\newcommand{\rr}{\mbox{${R}_{2}$}}
\newcommand{\rf}{\mbox{${R}_{1}$}}
\newcommand{\vtwo}{\mbox{${V}_{2}$}}
\newcommand{\vout}{\mbox{${V}_{1}$}}
\newcommand{\dshell}{\mbox{${{\rho}_{sh}}$}}
\newcommand{\dwind}{\mbox{${{\rho}_{w}}$}}
\newcommand{\dslow}{\mbox{${{\rho}_{s}}$}}
\newcommand{\dfast}{\mbox{${{\rho}_{f}}$}}
\newcommand{\vfast}{\mbox{${v}_{f}$}}
\newcommand{\vslow}{\mbox{${v}_{s}$}}
\newcommand{\cc}{\mbox{${\rm cm}^{-3}$}}
\newcommand{\apj}{\mbox{ApJ}}
\newcommand{\apjl}{\mbox{ApJL}}
\newcommand{\apjs}{\mbox{ApJS}}
\newcommand{\aj}{\mbox{AJ}}
\newcommand{\araa}{\mbox{ARAA}}
\newcommand{\nat}{\mbox{Nature}}
\newcommand{\aap}{\mbox{AA}}
\newcommand{\gca}{\mbox{GeCoA}}
\newcommand{\pasp}{\mbox{PASP}}
\newcommand{\mnras}{\mbox{MNRAS}}
\newcommand{\apss}{\mbox{ApSS}}

\date{}

\pagerange{\pageref{firstpage}--\pageref{lastpage}} \pubyear{2002}

\maketitle

\label{firstpage}

\begin{abstract}
SN1996cr is one of the five closest SNe to explode in the past 30
years. Due to its fortuitous location in the Circinus Galaxy at $\sim$
3.7 Mpc, there is a wealth of recently acquired and serendipitous
archival data available to piece together its evolution over the past
decade, including a recent 485 ks {\it Chandra} HETG spectrum. In
order to interpret this data, we have explored hydrodynamic
simulations, followed by computations of simulated spectra and light
curves under non-equilibrium ionization conditions, and directly
compared them to the observations. Our simulated spectra manage to fit
both the X-ray continuum and lines at 4 epochs satisfactorily, while
our computed light curves are in good agreement with additional
flux-monitoring data sets. These calculations allow us to infer the
nature and structure of the circumstellar medium, the evolution of the
SN shock wave, and the abundances of the ejecta and surrounding
medium.  The data imply that SN 1996cr exploded in a low-density
medium before interacting with a dense shell of material about 0.03pc
away from the progenitor star.  We speculate that the shell could be
due to the interaction of a blue supergiant or Wolf-Rayet wind with a
previously existing red supergiant (RSG) wind. The shock wave has now
exited the shell and is expanding in the medium exterior to it,
possibly the undisturbed continuation of the dense RSG wind. The
narrow lines that earned SN 1996cr its IIn designation possibly arise
from dense, shocked clumps in the CSM. Although the possibility for an
LBV progenitor for this Type IIn SN cannot be completely excluded, it
is inconsistent with much of the data. These calculations allow us to
probe the stellar mass loss in the very last phases ($ < 10^4$ years)
of a massive star's life ($> 10^6$ years) , and provide another means
to deducing the progenitor of the SN.
\end{abstract}

\begin{keywords}
circumstellar matter; methods: numerical; techniques: spectroscopic;
supernovae: individual: SN 1996cr; stars: winds, outflows; X-rays:
individual: SN 1996cr
\end{keywords}

\section{Introduction}
Core-collapse supernovae (SNe) arise from stars with zero-age
main-sequence masses $\ga 8 \msun$. However, decades of research has
failed to determine a direct relationship between the various SNe
types (as classified based on their optical spectra and light curves)
and the progenitor stars that gave rise to the SNe. It is not clear
how the type IIP, IIL, IIb, IIn, Ib and Ic types relate to the
properties of their respective progenitor stars. The progenitors
themselves, as it turns out, are not very well known, with the handful
of identified ones appearing to be associated almost exclusively with
Type IIP SNe \citep{smartt2009}. It had been surmised that the main
progenitors of core-collapse SNe were red supergiants (RSGs) and
Wolf-Rayet (W-R) stars \citep{falk1977, podsi92}. The explosion of SN
1987A revealed that blue supergiants (BSGs) could also be SN
progenitors \citep{sonnebornetal87}, perhaps in a binary system
\citep{morris2007, podsiadlowskietal2007}. In the last decade there
has been discussion of LBV stars being the progenitors of Type IIn SNe
\citep{chuetal99, salamanca00, kv06, vink2008, smith08, trundleetal08,
  trundleetal2009, galyam2009}, which is problematic because stellar
theorists have mainly placed the LBV stage as an intermediate
post-main sequence stage, not as a pre-explosion phase
\citep{schalleretal92, langer93, Langer1994, Stothers1996,
  garciasegura1996b, maederetal2005, maeder2008}.

The problem is clear - we rarely know the progenitor star that led to
a SN explosion, because it has to be typed from pre-explosion images,
leading to significant ambiguity and potential bias. The expansion of
the SN shock wave and the resulting emission due to circumstellar
interaction \citep{chevalier1994} opens up another window into the
exploration of the pre-SN star. The thermal emission from this
interaction, including the X-ray and optical emission, and to some
extent the non-thermal radio emission \citep{chevalier1982b}, depends
directly on the external density. Thus an accurate analysis and
interpretation of this emission acts as a probe of the density
profile. In the case of core-collapse SNe, which lose a significant
amount of mass prior to collapse, the surrounding medium is formed by
material from the pre-explosion star. Decoding the structure of this
circumstellar medium therefore will allow us to probe the mass-loss
parameters of the pre-SN star, which can then be linked to the stellar
parameters. Thus this provides a way to explore the stellar parameters
even after the star ceases to exist, and allows us to probe the pre-SN
properties of classes of SN that have heretofore lacked progenitor
counterparts.

In this paper we pursue this method for the unusual Type IIn SN
1996cr. This SN, which exploded around 1996 but was only discovered
about 11 years later, is only the second SN after SN 1987A which shows
increasing radio and X-ray emission over a sustained period of a few
years \citep{baueretal2008}. \citet{baueretal2008} suggested that the
increasing X-ray and radio emission may be associated with a dense
shell of material which the expanding SN shock wave interacts with a
couple of years after explosion. Herein we carry out hydrodynamical
simulations to evaluate this hypothesis, and to decipher the detailed
structure of the circumstellar medium into which the shock wave from
SN 1996cr is expanding. We compute the X-ray lightcurves and X-ray
spectra using non-equilibrium ionization conditions, which we compare
with high-resolution {\it Chandra} observations. We achieve a detailed
agreement which, given that we are using a single model across
multiple epochs, affirms the validity of our hydrodynamical model, and
allows us to constrain a range of abundances for both the material
ejected in the explosion and the surrounding circumstellar medium.

The plan of this paper is as follows: In \S \ref{sec:1996cr} we review
the observational details of SN 1996cr, including our recent High
Energy Transmission Grating (HETG) spectra which motivated this
analysis. \S \ref{sec:hydro} describes the reasoning behind our model
of the circumstellar medium (CSM), and the results of the
hydrodynamical modelling of the SN ejecta interacting with the
CSM. Our techniques for computation of the X-ray light curves and
spectra from the hydrodynamical models, and the resultant X-ray
emission, are outlined in \S \ref{sec:xray}. \S 5 puts the narrow
lines that earned SN 1996cr its Type IIn designation into the context
of the overall model. In \S \ref{sec:discussion}, the implications for
the progenitor, and the abundance determinations, are discussed in
depth. Finally, \S \ref{sec:summary} summarizes our results and
outlines future work in this area.

\section[]{SN 1996cr}
\label{sec:1996cr}
This luminous type IIn SN was only identified in the nearby Circinus
Galaxy \citep[3.7$\pm$0.3 Mpc;][]{koribalskietal2004} $\sim$11 years
after it was believed to have exploded \citep{Bauer2007,
  baueretal2008}, but its remarkable evolution was fortuitously
captured in archival data from {\it HST}, {\it Chandra}, {\it
  XMM-Newton}, {\it Spitzer}, and several ground-based optical and
radio observatories taken for other purposes. Initially undetected at
radio and X-ray wavelengths, the SN emission jumped by a factor of at
least $\sim$ 300 at radio wavelengths about 2 yrs after explosion. It
continued to increase modestly for 8 years before finally exhibiting
signs of a potential X-ray and radio turnover in late 2008.

\begin{figure*}
\includegraphics[angle=0,scale=1.]{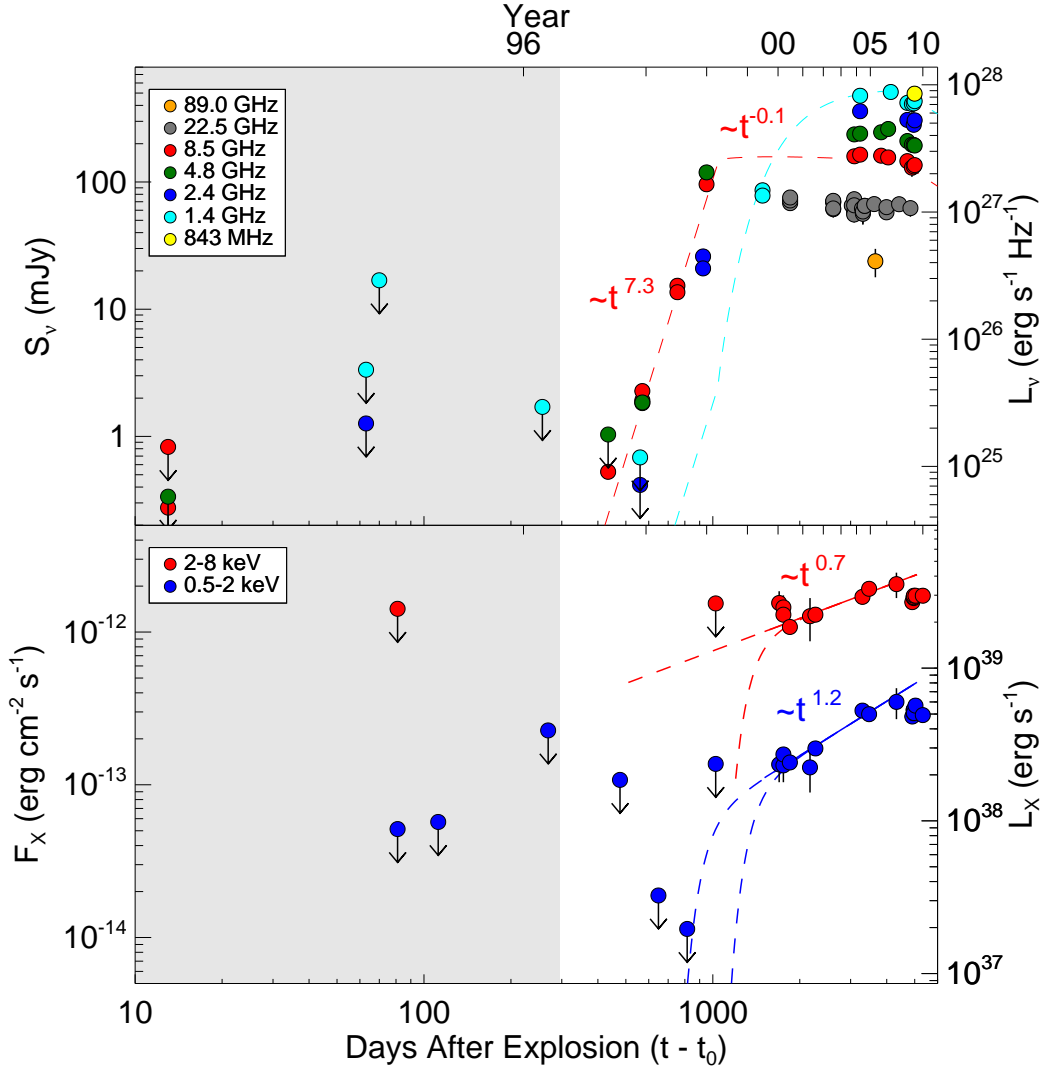}
\caption{(Top) The radio light curve of SN 1996cr. All available data
  from 843 MHz to 89 GHz are shown (some of which is previously
  unpublished and will be presented in Bauer et al., in
  preparation). Only upper limits are available during the first 3
  years. The radio flux clearly shows a sharp increase starting from
  around 700 days, a flattening of the light curve around 2000 days,
  and a decrease and turnover around 5000 days. On top of this
  luminosity evolution in the 8.5GHz band, we see a gradual decreasing
  absorption at lower frequencies (1.4GHz) that leads to a smoother
  transition between 1000-3000 days. (Bottom) The X-ray light curve of
  SN 1996cr over time. Unlike the radio, once detected the X-rays show
  a more-or-less constant increase over time, and no
  flattening. However the late-time turnover is present in the X-ray
  emission also. The upper limits in the first 3 years are quite high,
  with the points around 700 and 900 days being the most
  constraining. Sample power-law fits to the light curves are also
  shown. The simulations in this paper are geared towards explaining
  the X-ray emission, while being consistent with the radio
  behavior. \label{fig:96crlc}}
\end{figure*}

Fig.~\ref{fig:96crlc} shows the radio and X-ray evolution of SN
1996cr. Unfortunately, only upper limits exist for the first few years
after the explosion. But even these are sufficient to illustrate the
low level of emission in the initial years, followed by a steep rise
in the radio, and a less steep but no less significant rise in the
X-ray emission, which lasted for roughly a decade. This behavior is
reminiscent of the more famous SN 1987A, whose emission follows a
similar pattern \citep{mccray2003, mccray2007}, albeit with a factor
of $\sim$ 1000 lower luminosity \citep{baueretal2008}. Emission from
SN 1987A was explained by \citet{chevalier1995} as arising from the
interaction of the SN shock wave with a dense HII region interior to
the circumstellar ring seen in the equatorial region. Further
interaction with the denser ring should result in a continuously
increasing light curve in the case of SN 1987A \citep{Dwarkadas2007d}.

Early (2004) HETG X-ray data of SN 1996cr showed hints of the complex
velocity structure, similar to that seen in the optical. This, coupled
with the similarity with SN 1987A and the need to investigate such an
unusual SN further, motivated a deep 485ks {\it Chandra} HETG observation of
SN 1996cr in early 2009, with regular monitoring thereafter. A
companion paper (Bauer et al. 2010, in preparation) describes the
analysis of the HETG spectrum. In this paper we concentrate on the
interpretation of the X-ray emission, and the computation of the
structure of the medium into which the SN shock wave is expanding. The
steep rise in the X-ray and radio, and the more recent turnover in the
emission (see Fig.~\ref{fig:96crlc}) point to an interaction of the SN
shock wave with high density material, from which the shock wave has
now emerged. This possibility was already indicated by an analysis of
the radio spectrum earlier in \citet{baueretal2008}. Herein we provide
a more quantitative assessment of the various properties of the
circumstellar medium based on the X-ray data.

The 2009 HETG spectrum (shown later in Fig.~\ref{fig:96crspec}) is
probably one of the most detailed X-ray spectra of a young supernova,
next only to SN 1987A. The well-resolved emission lines allow us to
estimate the relative abundances of the ejecta to reasonable
accuracy. Combined with lower signal-to-noise HETG spectra from {\it
  Chandra} in 2000 and 2004, and a high signal-to-noise {\it
  XMM-Newton} p-n spectrum in 2001, as well as other CCD-quality X-ray
data, it provides a strong basis for accurate comparison of our
simulated spectra with the observations, allowing us to fine tune our
models.

Any model of the medium into which SN 1996cr is expanding must be able
to explain the increase in the radio and X-ray lightcurves, and the
spectral details. The model that we put forth in this paper is
constrained by the following observational parameters: (1) The range
of explosion dates, from March 1995 to March 1996
\citep[see][]{baueretal2008}. (2) Observed X-ray upper limits prior to
year 2000 and detected fluxes thereafter (Figure
\ref{fig:96crlc}). (3) A radio VLBI measurement that provides a size
measurement, which we take to be the radius of the outer shock. This
gives the radius of the outer shock as about 2.8 $\times 10^{17}$ cm
at 12 years (with a statistical error of $\approx$ 20\%).  (4)
Emission-line velocities seen in the X-ray and optical spectra,
ranging from 2000-5000 km s$^{-1}$. (5) Fits to the temperature (kT)
and the absorption column density at various times, obtained from the
X-ray spectra at several epochs. (6) Fits to prominent lines seen in
the emission spectra, which constrain CSM and ejecta abundances. (7)
The radio light curve (Figure \ref{fig:96crlc}).

\section{Ejecta and CSM Models and Hydrodynamical Calculations}
\label{sec:hydro}

In order to study the interaction of the SN ejecta with the
surrounding circumstellar medium, we need to delineate the density
profile of the ejecta as well as the surrounding medium. In this first
paper we make the assumption that the density structure, of both the
ejected material and the surrounding medium, is spherically symmetric,
and therefore that a one-dimensional (1D) model will suffice. Although
the complexity of line emission seen, especially in the optical
spectrum, may indicate otherwise, a spherically symmetric model is a
reasonable starting point, given the lack of any imaging
constraints. The validity of such an assumption can only be determined
by computation of the light curves and spectra resulting from these
calculations. As we shall show later, the spherically symmetric
assumption appears to be adequate in this regard. The companion paper
by Bauer et al (2010) will explore the line shapes and emission in
detail for deviations from spherical symmetry. This work will then
serve as a launching pad for more complicated explorations of the
density structure.

Following the work of \citet{chevalier1982a} and
\citet{chevalier1994}, we denote the ejecta density dropping off as a
power-law with radius (or velocity). \citet{matzner99} narrows this
power-law to a value of around 9 or 10; in the current work we adopt a
value of 9. In order to conserve mass and energy, the power-law
density profile cannot extend all the way to the origin, therefore
below a certain velocity the density is assumed to remain
constant. This ejecta profile model therefore has two free parameters
- the normalization of the outer density profile and the
transition-to-plateau velocity; alternatively, these two parameters
can be expressed in terms of the ejecta energy and mass.

Note that for early times, until the reverse shock reaches the plateau
region, only the single parameter of the density normalization is
involved in the hydrodynamics and there is a degeneracy between ejecta
mass and energy.  In this case one can set the energy to a canonical
value and use the ejecta mass as the free parameter.  Changing the
explosion energy scales this mass accordingly
\citep{chevalier1994}. Our late time modeling of SN 1996cr shows that
in order to match the light curve, the reverse shock must have passed
into the plateau region. Fitting the X-ray spectra constrains the
ejecta energy and mass to be 10$^{51}$ ergs and 4.45 $\msun$.  The
latter mass value is very reasonable for the total ejected material in
a core-collapse explosion \citep{Woosley2002}.  We note that this
gives a characteristic velocity $v_{\alpha} = 2 \times
E_{kin}/M_{ejecta} = 4740 $ km s$^{-1}$ which is suggestive of a
stripped envelope SN \citep{maureretal09}. In \S \ref{sec:prog} we
will discuss the connection with stripped envelope SNe in more detail.

\subsection{CSM Density Profile -Power-Law Model} Elucidation of the 
density structure of the CSM required considerable trial and
error. The simplest assumption of a CSM is one with a power-law
density profile r$^{-s}$, where $s=0$ for a constant density medium
and $s=2$ for a wind. A wind whose mass-loss parameters are changing
with time can result in other values of $s$. Therefore we first
explored the validity of such monotonic-with-radius models.

The light curve indicates that the X-ray emission began increasing
linearly about two years after explosion. In theory it is possible to
have the X-ray emission increasing with time even if the ambient
density is decreasing with radius. This can be shown in the context of
the self-similar model \citep{chevalier1982a} for SN evolution.
Detailed results for the X-ray emission using the self-similar model,
for various values of the parameters mentioned below, and
investigating several different assumptions for the forward and
reverse shocks, are given in \citet{flc96}. Herein we present a
simplified analysis that encapsulates the important ideas relevant to
this work.

The X-ray luminosity $L_x$ from a source depends on the electron density
$n_e$, the emitting volume $V$ and the cooling function $\Lambda$

\be
 L_x \sim {n_e}^2 \, \Lambda \, V
\ee

We assume that the density of the medium into which the SN shock wave
is expanding goes as $r^{-s}$. The emission arises from a thin shell
of radius $\Delta r$ at a mean radius $r$ , whose volume $V$ can be
expressed as $4 \pi r^2\, \Delta r$. Note that for self-similar
evolution, $\Delta r \propto r$, and therefore $V \propto r^3$. For a
SN in the early stages, the temperature is going to be much larger
than 10$^7$K, and the cooling function is assumed to vary as $T^{0.5}$
\citep{chevalier1994}. For a strong shock, $T \propto {v_s}^2$, and
therefore the cooling function $\Lambda \propto v_s \propto r/t$ in
the self-similar case.

Therefore we get 

\be 
L_x \sim\; r^{-2s} \,\frac{r}{t} \, r^3
\ee

which gives

\be 
\label{eq:lum}
L_x \sim \; \frac{r^{4 - 2s}}{t}
\ee

For s=2, a SN expanding in a wind with constant mass-loss parameters,
this gives the well-known result that the emission decreases inversely
with time, $L_x \propto t^{-1}$.

For s=1, we get $L_x \propto r^2 / t$. In the self-similar case, $r
\propto t^{\alpha}$, and therefore we get that $L_x \propto t^{2
  \alpha - 1}$. The parameter $\alpha = (n-3)/(n-1)$ for s=1 with
${\rho}_{SN} = A t^{-3} v^{-n}$ \citep{chevalier1982a}, which is
greater than 0.5 for all $n > 5$. Therefore $2 \alpha - 1 > 0$, and
the power-law exponent is always positive, leading to an X-ray
evolution that increases with time, even though the density is
decreasing as r$^{-1}$.

For s=0, $L_x \propto t^{4\alpha - 1}$. With $\alpha > 0.4$, (0.4
being the value it would have in the Sedov-Taylor stage), the X-ray
luminosity increases with time for all values of $\alpha$.

Theoretically, it is therefore possible to envision an increasing
X-ray luminosity with a CSM whose density is constant or even
decreasing with radius. The radius can then be expressed in terms of
the self-similar solution. We use the self-similar solution to show
that this approximation is not a good one in the current situation,
for the following reasons:

\begin{enumerate}
\item The radio spectra indicate that the absorption was high even two
  years after explosion \citep{baueretal2008}, and that as the SN
  shock expanded within this region the free-free absorption
  continually decreased. This suggests a region of high absorption
  into which the shock was expanding two years after the
  explosion. The absorption decreased as the shock made its way
  through this region. It would be difficult to explain this with a
  constant or continually decreasing density profile.
\item In the discussion above the post-shock temperature is related to
  the shock velocity, which varies as $v_s \propto t^{\alpha -
    1}$. Thus the temperature will vary as $T \propto t^{2\alpha -
    2}$. For example, in order to get an X-ray luminosity $L_x \propto
  t^{1.0}$, roughly intermediate between the hard and soft luminosity
  slopes seen in Figure ~\ref{fig:96crlc} , we can take $s = 0.6$ and
  $n=9$. Then $\alpha=0.7142$, and $L_x \propto t^{1.0}$ from equation
  \ref{eq:lum}. This would mean that the temperature would vary as
  t$^{-0.57}$, which equates to an expected temperature drop of ~1.7
  between 2000 and 2009. Such a temperature variation is not
  seen. Instead the X-ray data (Bauer et al. 2010) above 2 keV show
  little spectral variation, and hence little apparent temperature
  evolution, over 9 years. The spectrum below 2 keV does vary
  somewhat, although it is not clear whether this is because of
  decreasing absorption in the soft X-ray regime or whether it is
  because the spectrum is actually becoming softer.
\item A radius of 2.8e17 at a time of 12 years implies an average
  velocity of 7500 km s$^{-1}$.  The optical and X-ray spectra do not
  show velocities exceeding 5000 km s$^{-1}$, and in most cases
  somewhat lower. If we assume spherical symmetry and an average
  velocity of 4000 km s$^{-1}$ from 2000 to 2008 (when spectra are
  available) then it suggests that the average velocity in the first
  4-5 years (depending on explosion date) must have exceeded an
  average velocity of 12,000 km s$^{-1}$. A power law model has shock
  velocity varying as $v_s \propto t^{\alpha - 1}$, which would make
  it extremely difficult, if not impossible, to fit such a velocity
  profile for reasonable values of $\alpha$. Whereas a model with a
  lower CSM density early on and a higher density later could fit it
  easily. Using our earlier example, we can construct a model with
  $n=9$ and $s=0.6$ using the self-similar solution, which satisfies
  the constraint that the shock radius be 2.8e17 at 12 years. This
  model gives a velocity of 5290 km s$^{-1}$ at 12 years. While this
  is somewhat high, it could be considered acceptable within the error
  bounds. However such a model would also predict a velocity of almost
  6000 km s$^{-1}$ at 8 years and 6450 km s$^{-1}$ at 6 years, which
  is not supported by the observations. Note that a constant density
  model with n=9 would have $v_s \propto t^{-0.33}$, which would make
  this discrepancy worse.
\item In a model with monotonically decreasing density, the density of
  the CSM would need to be significantly high throughout the evolution
  for the shock velocity to decelerate to a few thousand km s$^{-1}$
  in a few years, as suggested by the X-ray and optical spectra.  For
  instance, in the above model with $n=9$ and $s=0.6$, the density at
  5 $\times 10^{16}$ cm would be 7.71e-21 g cm$^{-3}$. This is about
  3800 times greater than what is in our simulations presented in the
  next section, which would result in the X-ray emission, especially
  the hard X-rays, being about 10$^7$ times greater on average,
  significantly exceeding the upper limits derived from archival
  observations.
\item The radio emission is increasing significantly with time after
  about 2 years (Figure \ref{fig:96crlc}), which is hard to accomplish
  in a model with decreasing density.  In particular free-free
  absorption by itself cannot account for this increase
  \citep{baueretal2008}, there still needs to be a flux increase $>
  100$. This can best be implemented with a density increase, as the
  radio emission, although non-thermal, indirectly depends on the
  external density \citep{Chevalier1996}.
\item The X-ray emission increases significantly after two years.  A
  simple power-law increase in the X-ray flux, as predicted by the
  self-similar model, cannot match the data, which requires at least
  two distinct power-laws, with the upward slope becoming much steeper
  after 2 years. This would then require at least two different
  density slopes, with the density decreasing slower in the second
  case. When combined with the VLBI radius constraint, and the
  velocity and temperature constraints outlined above, it would be
  almost impossible to make the power-law model consistent with all
  the available data.
  
\end{enumerate}

The above arguments suggest that a CSM density profile that shows a
simple power-law decrease would not work. Although a self-similar
solution is not applicable, many of the same arguments can be used to
show that an increasing power-law density profile from the origin (the
stellar surface) also would not work. Instead, a density profile where
the density changes after 2 years is suggested.

\subsection{CSM Density Profile - High Density Shell} The above discussion 
clearly argues for these basic requirements of any CSM model for SN
1996cr: a CSM whose density is low close to the star, then increases
sharply with radius. This would fulfill both the velocity and density
criteria outlined above, as well as lead naturally to the observed
radio spectra and light curves. One way to arrive at such a
distribution is with a low density wind medium followed by a thin,
dense shell, as is often seen around massive stars \citep{Weaver1977,
  chu03, cappaetal03, chu08}, and in the nebula around SN 1987A
\citep{bl93}.  In this paper we consider such a distribution. The
inner and outer radius of this shell, the shell density, and
refinement of the CSM density structure, are obtained via an iterative
procedure that consisted of adopting a value for the shell parameters,
computing the hydrodynamic interaction, calculating the X-ray light
curves and spectra, and re-iterating until a reasonable fit was
obtained between the simulated and observed spectra and
light-curves. This took on the order of a dozen iterations before
convergence was obtained.

\begin{figure*}
\includegraphics[angle=270,scale=.75]{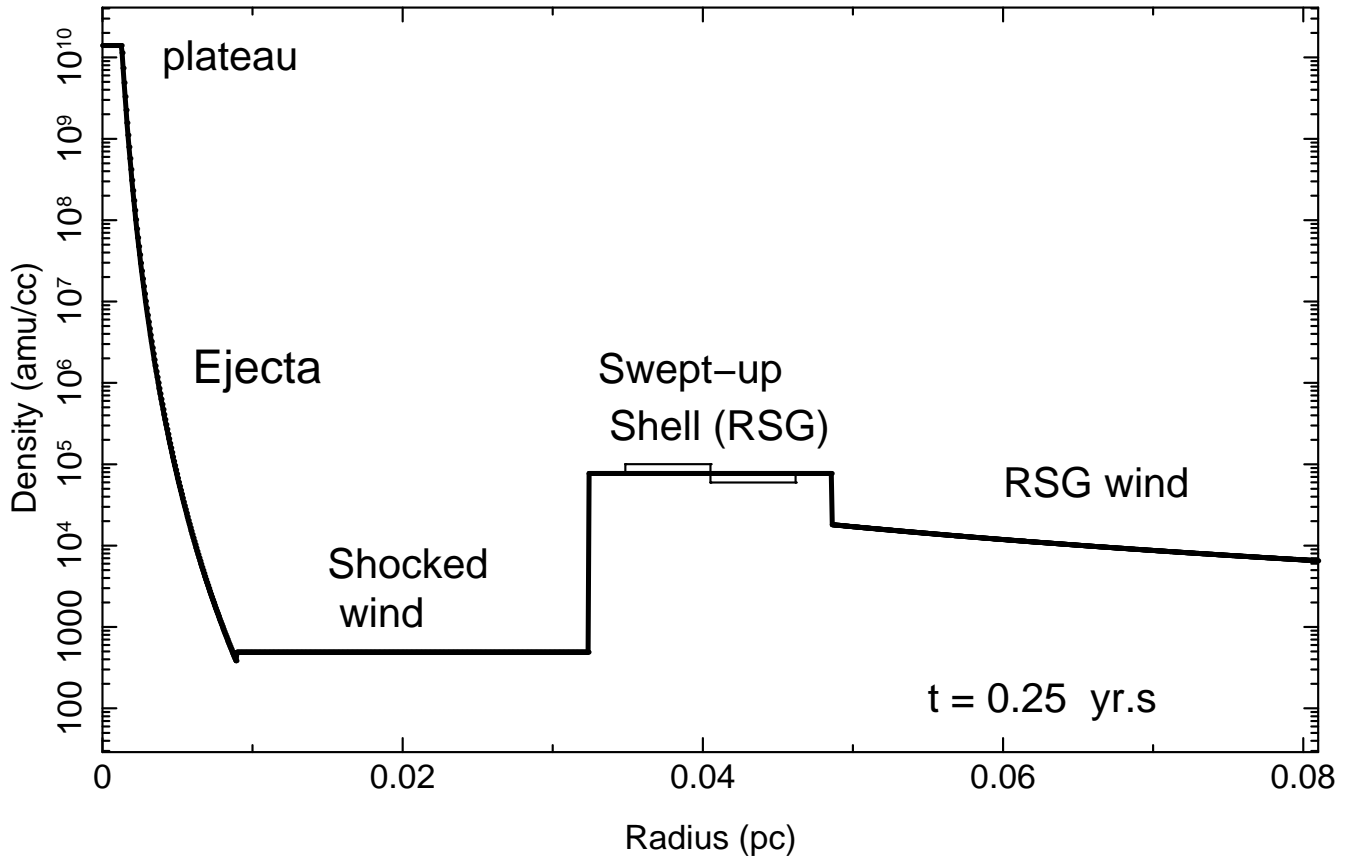}
\caption{The initial density profile of the SN ejecta and CSM, used as
  initial conditions in the hydrodynamic simulations.  Starting from
  left and moving outwards in radius, the density profile shows the
  constant density plateau region of the SN ejecta, followed by the
  steeply decreasing ejecta profile which goes as r$^{-9}$.  Beyond
  this is the circumstellar medium. The ejecta have already crossed
  the freely expanding wind region (not shown). Going outwards in
  radius, we encounter a uniform density shocked-wind region, the high
  density shell, and the freely expanding outer wind. A suitable
  temperature was assigned to each component. The pressure (and
  temperature) that we use for the initial conditions is not
  important, since it is significantly smaller than the pressure of
  the gas shocked by the SN blast wave.  The labels attributed to the
  various sections of the CSM reflect the discussion in \S
  \ref{sec:discussion}. The thin solid is intended to show how a
  rearrangement of the mass in the shell in this fashion, without
  changing any other parameters, can lead to a modest increase in the
  light curve at early times ($<$ 3 years), shown as the dotted line
  in Figure \ref{fig:xrayflux}.  \label{fig:hydroinit}}
\end{figure*}

The model that we finally converged on has a shell with density around
1.28 $\times 10^{-19}$ g cm$^{-3}$ extending from 1 $\times 10^{17}$
to 1.5 $\times 10^{17}$ cm. The X-ray light curves and spectra are
highly sensitive to the shell parameters, i.e.  the inner and outer
shell radii and shell density, and so manage to constrain them quite
tightly. Even a 20\% deviation of the shell radii from these values
destroys the agreement with the lightcurve, while a factor of 1.3 in
the density has about the same effect. Therefore the shell itself is
strongly constrained by the X-ray data. A rearrangement of the mass in
the shell, i.e. a modest change in the density profile that does not
alter the total mass (Figure \ref{fig:hydroinit}), can result in
modest changes to the early-time X-ray flux, as shown in Figure
\ref{fig:xrayflux}. For the CSM {\it inside} of the shell, we have
assumed a structure resemblant of a wind-blown bubble, with an initial
freely expanding wind whose density drops as r$^{-2}$ ending in a wind
termination shock, which is assumed to be a strong shock. Beyond this
is a constant density shocked wind region. The wind density, and
location of the wind termination shock, are constrained only by the
density of the shocked wind region, and are obtained as described in
the Appendix. For the standard case described here and shown in Figure
\ref{fig:hydroinit}, the value of the wind parameter $\mdot/v_w$ is
6.76 $\times 10^{11}\, {\rm g}\; {\rm cm}^{-1}$, the wind termination
shock is located at 1.63 $\times 10^{16}$ cm, and the shocked wind
region has a constant density of 8.15 $\times 10^{-22}\, {\rm g}\;
{\rm cm}^{-3}$.

The structure of the medium {\it beyond} the dense shell is difficult
to estimate properly, since the shock front has not had much time to
expand within it. One assumption is that the dense shell consists of
fully swept-up wind material. In this case the CSM density structure
resembles a wind-blown bubble formed by the interaction of a
supersonic wind with the slower wind from a previous epoch; the dense
shell is composed entirely of the slower wind material that has been
swept-up. The mass of the shell must equal the total wind mass up to
the outer radius of the shell, or $M_{shell} = ({\dot M}_{w}/V_w) \;
R_{shell}$, where the subscript $w$ refers to the outer wind. Since
the shell outer radius is known, and the mass is defined once we have
the shell boundaries and shell density, this provides us with the wind
parameters ${\dot M}_{w}/V_w $. Assuming constant wind parameters, the
density of the material outside the shell at any given radius $r$ can
then be expressed as that of a freely expanding wind $\rho = {\dot
  M}_{w}/V_w \times 1/(4 \pi r^2)$. This was one of the two
assumptions that we have tried, and since it gives satisfactory
results over the short period of time that the shock is interacting
with this region, we have continued to use it. (The other assumption
involved a constant density profile exterior to the shell. However in
that case the shell mass was always found to be larger than the mass
of the swept-up material given that density, suggesting a complicated
origin for the shell. A constant density profile either gives too
small a density close in or too large a density further out, depending
on its magnitude. Furthermore, the total absorption column from any
constant density profile that provides adequate X-ray emission beyond
the dense shell increases too fast, and thus such a profile cannot be
sustained beyond 0.4pc in any reasonable model.  Since the wind
profile anyway fits better, we did not explore the constant density
assumption further).

The final density profile for the SN ejecta and CSM that we have used,
which forms the initial conditions for the hydrodynamical simulations,
is shown in Fig.~\ref{fig:hydroinit}. The thin solid lines
superimposed on the constant density shell are intended to show how
rearranging the mass to perturb the density in this fashion can lead
to a modest change in the X-ray lightcurve (the dashed gray line in
Figure \ref{fig:xrayflux}).

\subsection{Hydrodynamic Simulations} We carried out 
hydrodynamical simulations to compute the interaction of the SN ejecta
with the CSM. The simulations were carried out using VH-1, a
three-dimensional finite-difference hydrodynamic code based on the
Piecewise Parabolic Method of \citet{cw84}. Although cooling is
included in these simulations via the inclusion of a cooling function,
it was not found to play an important role in the 1D
calculations. However the cooling time of the shock within the shell
was of the same order of magnitude as the evolution time, and it is
possible that in multi-dimensional calculations, where the growth of
instabilities can give rise to dense clumps, cooling could prove to be
more effective\footnote{Our X-ray spectral calculation, as well as the
  formation of narrow optical lines, does in fact require the presence
  of dense clumps, but they comprise at most 1\% of the mass.  In
  these scenarios cooling is important, and in fact the shock within
  the clumps is assumed to be radiative in order to account for the
  narrow-line emission. However this would affect only a small portion
  of the material. In any case, these clumps are not included in the
  spherically symmetric numerical hydrodynamics simulations, and they
  would not affect the general hydrodynamical evolution described
  herein.}. The simulation was started with the initial conditions
described above, and the SN ejecta allowed to expand into the
surrounding medium.

\begin{figure*}
\includegraphics[angle=0,scale=.7]{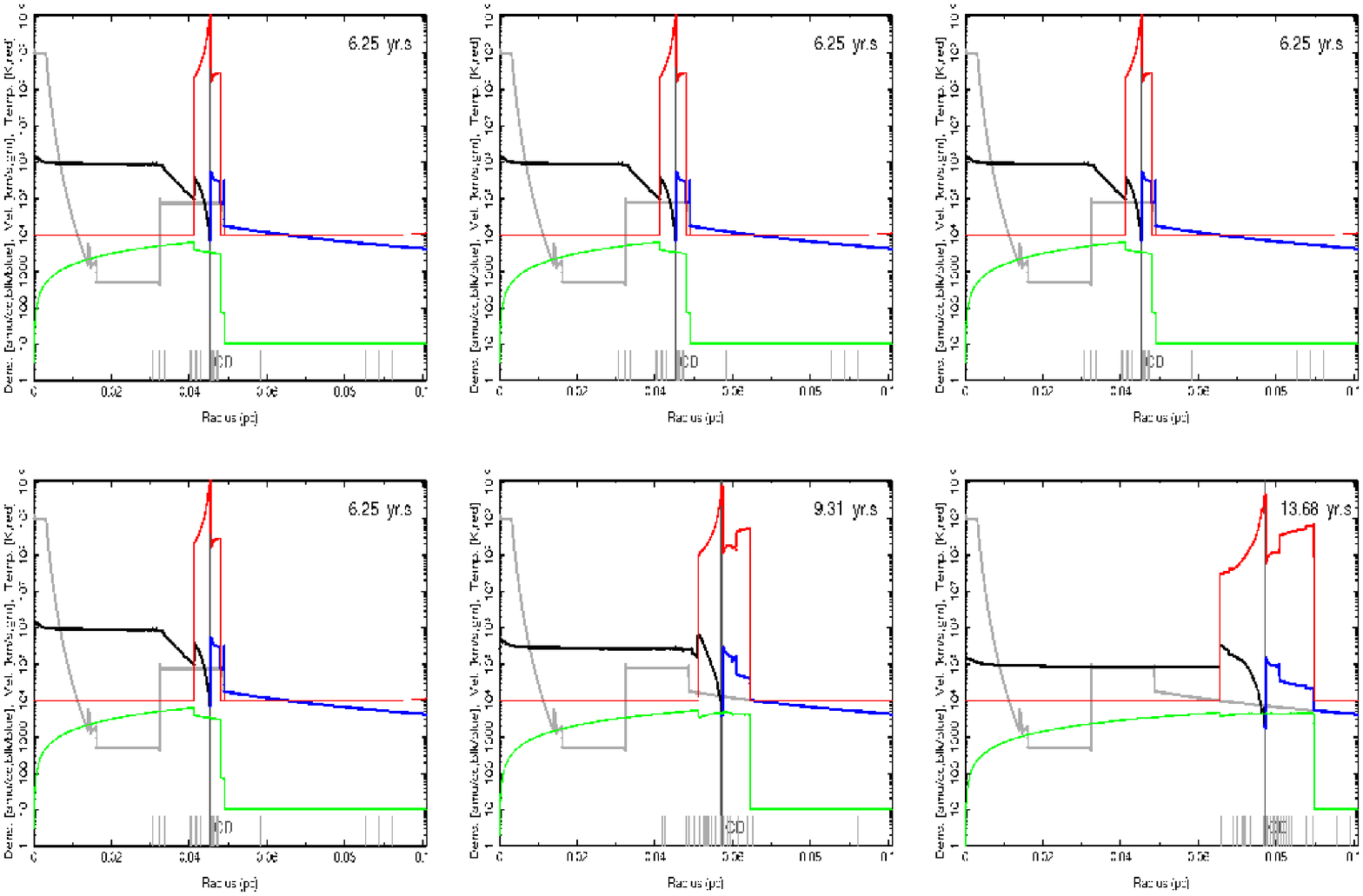}
\caption{These panels depict the hydrodynamic evolution of various
  quantities: ejecta density (black); CSM density (blue); temperature
  (red); and fluid velocity (green). The grey vertical lines denote
  the borders of the mass-shells into which the domain is divided up,
  for the purpose of computing and tracking the fluid history. The
  thick grey line shows the initial density profile to enable the
  reader to see how it changes with time. The last four panels
  approximate the times when the spectra outlined in \S 4 were
  taken. The various panels show, from left to right and top to bottom
  (1) the initial interaction of the SN ejecta with the shocked wind,
  and the formation of a forward and reverse shock (2) The interaction
  with the dense shell, the compression of the interaction region
  between the shocks, and the formation of a resultant transmitted and
  reflected shock (3) (HETG-00) The transmitted shock is advancing
  slowly through the shell, while the reflected shock goes up the
  steep ejecta incline. Note that the interaction region is slowly
  growing bigger. (4) (XMM-01) The forward shock has reached the edge
  of the shell, and will soon start to expand out into the external
  wind. (5) (HETG-04) The reflected shock reaches the constant density
  plateau region of the ejecta. Beyond this point, the solution is no
  longer self-similar, if it indeed ever was given the various
  transitions. (6) (HETG-09) The evolution of the forward shock in the
  external wind, while the reverse shock is expanding back into the
  ejecta plateau region.  \label{fig:hydroevol} }
\end{figure*}

\begin{figure*}
\includegraphics[angle=0,scale=.75]{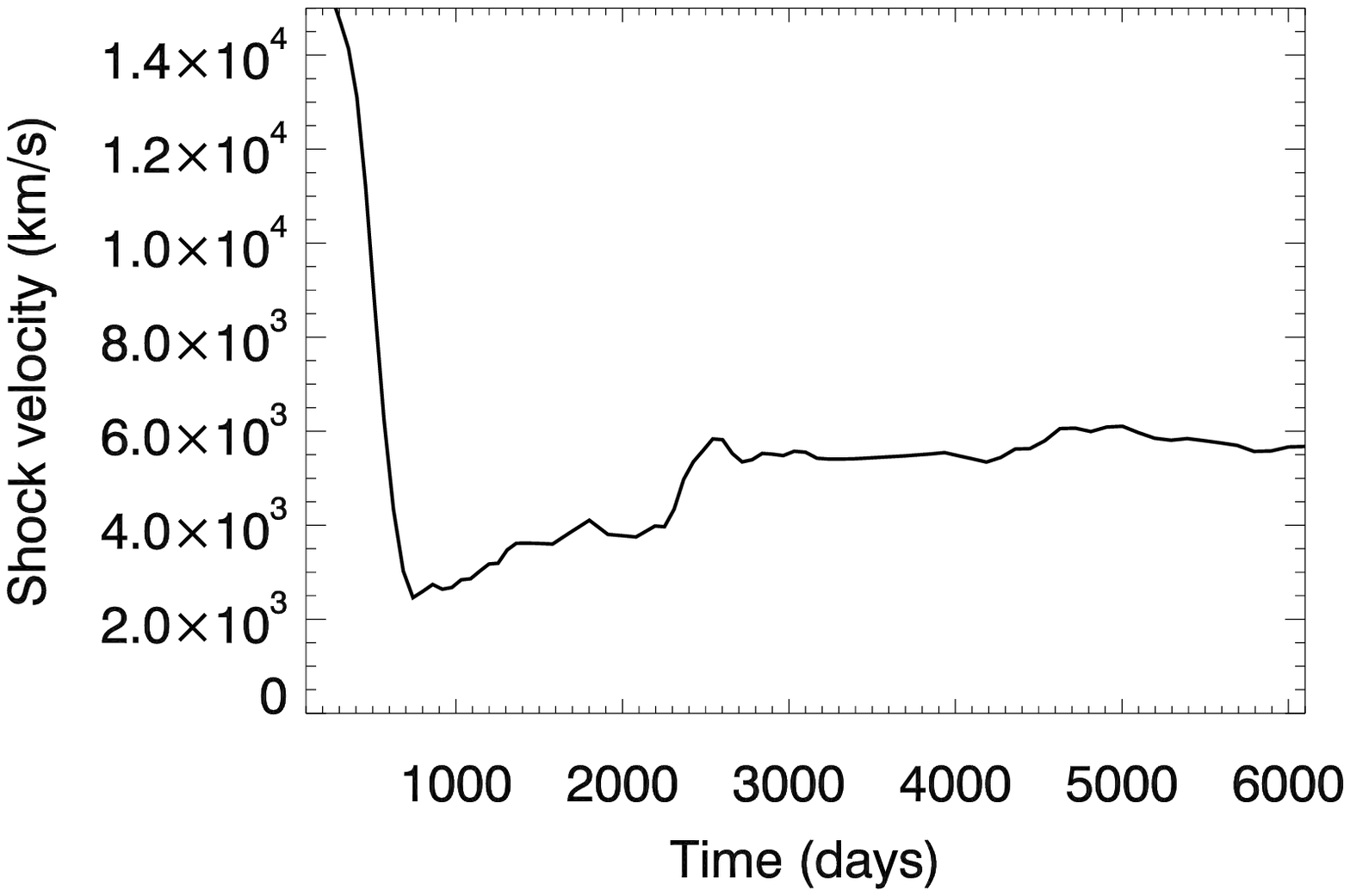}
\caption{The velocity of the SN forward shock during the evolution. It
  is extremely high (exceeding 20,0000 km/s) in the first few months,
  but then drops dramatically when the shock encounters the higher
  density shell, and the interaction region is highly compressed. As
  the compressed region slowly expands, the velocity rises again as
  the built-up pressure is released (A similar effect was observed in
  SN 1987A). After about 6 years, the shock exits the shell, and the
  velocity increases as it encounters a lower-density medium. As the
  density profile of the shocked interaction region changes from that
  in a constant-density medium to that in a wind medium, the velocity
  remains more-or-less constant, at a velocity around 5500 km
  s$^{-1}$.  \label{fig:shkvel} }
\end{figure*}

The evolution of the various dynamical and kinematic quantities over
time are shown in Fig.~\ref{fig:hydroevol}. The SN ejecta expand
quickly in the low density medium (see Fig.~\ref{fig:shkvel}) until
they reach the higher density shell in about 20 months. The shock,
which was expanding very rapidly up to this time, is instantaneously
decelerated by the high density, and its velocity drops to about 2000
km s$^{-1}$ due to the high pressure. As seen in
Fig.~\ref{fig:hydroevol}, the high pressure region between the forward
and reverse shocks gets highly compressed. The collision of the ejecta
with the high density shell sends a lower velocity transmitted shock
into the dense shell, and a reflected shock is formed that travels
back into the ejecta. The reflected shock quickly overtakes the
original reverse shock, thus thermalizing the ejecta faster than the
SN shock itself would have done. The transmitted shock gradually
increases in velocity as the highly over-pressured region slowly
depressurizes, and then attains a more or less uniform velocity of
around 5500 km s$^{-1}$.

The interaction of the shock wave with the high density shell lasts
for about 4.6 years (from t $\sim$ 1.5 to t $\sim$ 6.1 years.). The
structure of the interaction region resembles the interaction of a
power-law profile with a constant density medium, as expected, and the
density decreases from the reflected shock to the contact
discontinuity. Since the pressure does not vary much, the temperature
is almost inversely proportional to the density, and increases
outwards from the reflected shock, until it reaches a large value at
the contact discontinuity. We note that the structure is comparable to
that described by the self-similar solution for a steep ejecta density
profile colliding with a wall \citep{Chevalier1989}. The comparison is
close but not exact, since the shock in this case imparts some motion
to the shell, which is not included in \citet{Chevalier1989}.  It is
also interesting to note that, given the presence of the dense shell
so close to the star, the density into which the forward shock is
propagating, while in the shell, is much higher than the density into
which the reverse shock is propagating. This is contrary to a general
expansion in a power-law or constant density medium, and results in
the temperature behind the forward shock ($\sim 10^8$) being initially
{\em lower} than the temperature behind the reverse shock ($\sim
10^9$). Therefore we expect that the major contribution to the X-ray
emission in the 0.5-10 KeV wavebands at this early time arises from
the shocked circumstellar material. As the shock continues to expand,
the temperature behind the forward shock increases, whereas that
behind the reverse shock decreases as the shock expands into the
increasing steep power-law ejecta density profile. When the
transmitted shock exits the shell and begins to reform in the freely
expanding wind around it, the reflected shock is then interacting with
the constant density ejecta, whose density is higher than that of the
wind medium. The density behind the reflected shock is higher than
that behind the forward shock, its temperature lower, and the emission
becomes dominated by the reverse shocked ejecta. It was found that a
good fit to the late-time emission required that the reverse shock
enter the plateau ejecta region.  The location of the plateau, the
velocity below which the ejecta density remains constant, was adjusted
to match the observations, fixing the total ejected mass to the value
given in the previous section.

Once the transmitted shock exits the shell, it is expanding in the
lower density wind, whose density decreases as r$^{-2}$. The radius of
the forward shock at $\approx$ 12 years is consistent with that
obtained from VLBI imaging to within 5\%. Thus this approximation has
proved suitable so far, but only time and future data can confirm
whether this approximation is valid for the further expansion of the
shock wave. While most of the X-ray emission at this point is expected
to arrive from the reverse shocked ejecta, the radius and velocity of
the forward shock, and therefore the size of the remnant, are
determined by the interaction with this wind.

It is interesting to note that although the shock is interacting with
an r$^{-2}$ wind, the shock structure is resemblant of the shock
interacting with a constant density medium. This is to be expected as
it will take a few dynamical times for the shock to reach the
self-similar solution in the wind \citep{Dwarkadas2005}. Over time
therefore the entire density structure of the shocked interaction
region is predicted to change, with the density at the contact
discontinuity going from relatively small to relatively large
values. Such a change should be reflected in the X-ray emission over
the next few decades.

\section{Computation of the X-ray Emission and Comparison with Observations}
\label{sec:xray}
\subsection{Non-equilibrium Ionization and Lagrangian Shells}

Our goal is to compute the X-ray emission from the output of the
hydrodynamic simulation efficiently and with reasonable accuracy.
Plasma that has been recently shocked and whose density is low is
presumably not in ionization equilibrium \citep{hamiltonetal1983,
  borkowski2000, borkowskietal2001}; as a rule of thumb this is the
case if $\tau \equiv n_e t < 10^{12}$~s\,cm$^{-3}$ where $n_e$ is the
electron density and $t$ is the time since the plasma was shocked.
Since the plasma impacted by the SN shock wave has a density typically
less than 10$^5$~cm$^{-3}$, the time to reach ionization equilibrium
is measured in years and the X-ray emission must be computed using
non-equilibrium ionization (NEI) conditions.  In general this is done
by calculating the evolution of ion fractions using ionization and
recombination rates under appropriate simplifying assumptions
\citep{hamiltonetal1983, borkowskietal1994}.  Here, however, our
approach is to make use of existing NEI codes and appropriately
interface them to the hydrodynamics output.

We use NEI emission models available in the XSPEC package, {\tt vgnei}
and {\tt vpshock}, along with ``NEI version 2.0'' atomic data from
{\tt ATOMDB} \citep{smith2001} which has been updated to include
relevant inner-shell processes\footnote{Files provided by K. Borkowski
  and online at {\tt http://space.mit.edu/home/dd/Borkowski/}}.  These
codes encode the NEI state with only a few parameters; we have chosen
to sacrifice some accuracy for speed and simplicity.  To calculate the
X-ray emission from an individual fluid element these models need
information that is based on the element's history, in particular the
ionization age, $\tau = \int n_e(t)\, dt$, the current electron
temperature, $T_e$, and the ionization-age-averaged temperature,
$<kT_e> = \int T_e(t) n_e(t)\, dt / \tau$ \citep{borkowskietal2001}.
These quantities can be calculated from the set of simulation time
steps provided the location of a given fluid element can be determined
and followed in time.  To this end we regroup the Eulerian output at
each time step in a Lagrangian manner, using cumulative mass as the
Lagrangian parameter to define and track a set of mass intervals,
representing ``shells'' in 3-D.  These mass shells, of order 50 of
them, are not as finely spaced as the hydro radial bins (thousands),
so we calculate shell-averaged hydrodynamic fluid parameters: pressure
$p_j$, density $\rho_j$, and radial velocity $v_j$, for each shell at
the time steps, $t_j$.

\subsection{Abundances, $\mu$, and $T_e$}
\label{sec:abundseq}

The hydrodynamic variables are converted to values relevant for the
X-ray emission, $n_e$, $T_e$, through the abundances and the related
mean mass per particle, $\mu$.  We assign an abundance to each shell
of material: shells interior to the contact discontinuity are assigned
ejecta abundances and those exterior are CSM. It is possible (and as
we argue in the discussion section, most likely) that the CSM can be
divided into two phases, the portion interior to the shell and the
portion including the shell and exterior to it. However the evolution
of the SN ejecta into the region interior to the shell is not
constrained by the data. Therefore a single set of abundances is
assumed for the CSM for simplicity. It is likely that the ejecta may
be layered, as is expected for a massive star, and different abundance
distributions should be expected for different layers. However, we
have no {\it a priori} indication of any of the abundance
distribution, and 4 spectra of different resolution spread over 9
years, only the last two of which are actually tracking
reverse-shocked emission according to our calculations, would be
unable to distinguish between abundances of various layers. Therefore
as a first approximation we have considered a single set of ejecta
abundances, although our code is easily capable of assigning a unique
abundance to each shell of material if required.

Define $X(Z)$ as the relative abundance values with respect to a
reference number-ratio abundance set, A$_{AG89}$(Z); here we use the
\citet{ag89} abundance set.  The mean atomic weight can be calculated
as: \be \mu_A = \frac{\sum_Z m_Z\, X(Z)\, A_{AG89}(Z)}{\sum_Z X(Z)\,
  A_{AG89}(Z)} \ee where $m_Z$ is the mass of an ion of element $Z$.
Similarly the mean charge state averaged over all ions is given by:
\be q_A = \frac{\sum_Z Q_Z(T_{\rm CIE})\, X(Z)\, A_{AG89}(Z)}{\sum_Z
  X(Z)\, A_{AG89}(Z)} \ee where $Q_Z(T_{\rm CIE})$ gives the average
charge state of ion $Z$, i.e., the average number of free electrons
per $Z$-ion, and T$_{\rm CIE}$ refers to the temperature in
collisional ionization equilibrium.  This value changes with
progressive ionization, however it is dominated by the low-$Z$
elements which are quickly ionized, so we approximate its value based
on ionization equilibrium at a fixed temperature, $T_{\rm
  CIE}=10^7$\,K.

The mean mass per particle is then: \be \mu = \frac{\mu_A}{1+q_A} \ee
where $\mu$ here has units of mass.\,\footnote{Note that it is common
  to informally express $\mu$ in implied atomic mass units,
  $\mu/m_{\rm amu}$, e.g., ``$\mu$ = 0.5 for a fully ionized H
  plasma''.}  The densities of electrons and ions can now be obtained
from the fluid density as: $n_e = q_A\, \rho/\mu_A$, $n_i =
\rho/\mu_A$ and for a specific ion: $n_Z = n_i X(Z)\, A_{AG89}(Z)\, /
\sum_Z X(Z)\, A_{AG89}(Z)$.

We can now compute the average particle temperature in an element of
plasma from the hydrodynamic pressure and density: $kT_{\rm hydro} =
\mu p/\rho$.  Since the shocks are collisionless, most of the
post-shock energy is transferred to ions, and the electron temperature
is generally a fraction $\beta$ of the ion temperature
\citep{ghavamianetal2007}.  This is accounted for by setting $T_e =
\beta_{\rm sh}(v_{\rm sh})~T_{\rm hydro}$ with the $\beta_{\rm sh}$
value set to be $\propto v_{sh}^{-2}$ as given in
\citet{ghavamianetal2007}; the shock velocity is calculated
self-consistently from $kT_{\rm hydro} = (3/16) \mu ~ {v_{\rm sh}}^2$.
Other choices for $\beta_{\rm sh}(v_{\rm sh})$ can be easily
implemented if desired.  For example, we clipped $\beta$ to have a
minimum value of 0.05 as an approximation to the results of
\citet{vanadelsbergetal08}; however, this produced no significant
changes in our calculated flux or spectra as the $\beta$ values of the
mass-cuts are generally well above 0.1.

Finally, the $\beta$ value is modified to include its evolution due to
Coulomb interactions: $T_e/T_{\rm hydro}$ approaches 1.0 as the
ionization age approaches $\tau_{\rm equil}(v_{\rm sh})$.  The
dependence on shock velocity is given roughly by: $ \tau_{\rm equil}
\sim 10^{10} \, (v_{\rm sh}/v_{400})^3 $, with $v_{400}$ = 4$\times
10^7$~cm\,s$^{-1}$; this follows from the $T^{3/2}$ variation of the
Coulomb equilibration time constant. For smaller values of $\tau$,
$\beta$ changes such that $ \beta \sim (\tau/\tau_{\rm equil})^{0.44}
$, e.g., as seen in Figure~4 of \citet{michaeletal02}.  With these
relations the value of $\beta$ is increased from its initial
post-shock value when the $\tau$ of the mass-cut increases
sufficiently.

\subsection{Implementation and Embellishments}

Using the above equations, for each shell at each time step we can
calculate the input values needed for the {\tt vgnei} model
\citep{borkowskietal2001}: {\tt kT}, {\tt Tau}, {\tt <kT>}, and the
      {\tt norm}\,$=~10^{-14}\,\Delta V\,n_en_H\,/(4\pi d^2)$, with
      $\Delta V$ the shell volume and $d$ the source distance.  The
      {\tt vgnei} abundance values are the appropriate set (ejecta,
      CSM) scaled so that the H abundance is 1.  The calculations,
      from reading the hydro output files to evaluating the XSPEC
      spectral model, are all done with custom routines written in
      S-Lang, the interactive scripting language used in the ISIS
      package \citep{houck2000}.

The flexibility of the code also allows us to implement additional
modifications to improve the fidelity of the emission calculation as
needed.  For most mass shells we use the {\tt vgnei} model with an
appropriate $\tau$ value based on the shell history as described
above.  For the case where a mass shell is partially shocked, the
shock front is within the shell and a constant $\tau$ value is not
appropriate since there is a range of $\tau$ starting at $\tau=0$ up
to a maximum value.  In this case we can calculate the shell's
emission using the {\tt vpshock} model, which allows us to set a range
of $\tau$ from a low (zero) to a high value. Inclusion of these
partially-shocked shells adds some line emission of lower ion states
and smoothes the light curves.

There are indications that the X-ray emission from SN 1996cr is not
simply the sum of optically thin components: the as-fit $N_H$
decreases with observation epoch and the X-ray line shapes at
high-resolution are asymmetric.  These data are examined in more
detail in the companion paper (Bauer et al.~2010, in preparation) but
the methods to include them in the emission model are briefly
summarized here.  To self-consistently model the $N_H$ effect, we can
add ``internal absorption'' proportional to the amount of unshocked
CSM material external to the forward shock.  In the case of the
asymmetric line shapes it seems likely, and our hydro model predicts,
that there is substantial opacity through the SN ejecta core.  Hence
some of the red-side emission is obscured producing asymmetric lines
and reducing the observed flux.  Based on simple 3D modeling
\citep{dn09}, a value of 0.75 has been used to account for the amount
of flux typically obscured by the core, and a blue-shift and Doppler
Gaussian broadening, proportional to the hydro shell velocity, are
applied as a first approximation to the line shape from the shell.

\subsection{Comparison with the Data}
\label{sec:datacomp}

As the previous sections imply, the X-ray emission from a given hydro
simulation depends on some additional parameters and assumptions.
Perhaps the most influential of these are the elemental abundances
assigned to the hydro plasmas. These abundances show up in two main
ways: i) they determine the $\mu$ value of the plasma and hence affect
the electron temperature and the number of electrons per ion (section
\ref{sec:abundseq}), and ii) the abundances of the higher Z elements
are responsible for the X-ray emission lines in the spectra. Because
we use the XSPEC {\tt vgnei} ({\tt vpshock}) routine, the abundances
of the 13 elements currently treated by that model are the ones
relevant here: H, He, C, N, O, Ne, Mg, Si, S, Ar, Ca, Fe, Ni. These
abundances are specified relative to the \citet{ag89} values.

The amount of the lowest-Z elements, especially the ratio of H to He,
is most responsible for determining the mean mass per particle $\mu$
and thus changing the electron temperature proportionally.
Temperature changes lead to changes in the continuum shape and in the
ratios of He-like to H-like line emission from higher-Z elements.
Because their line emission is not clearly visible here, yet they
contribute most of the X-ray continuum, we refer to the elements H,
He, C, N, and O as ``continuum elements".  Using only X-ray data it is
hard to decide among different possibilities for these ``continuum"
abundances: H and He produce no lines in the X-ray range, and
depending on the temperatures and amount of absorption involved, it
can also be difficult to see lines to constrain C, N and even O
abundances with X-ray spectra.  For this reason we have little handle
on the abundances of C, N, and O in SN 1996cr.

With the continuum abundances set, the abundances of the higher-Z
elements can generally be varied quite a bit without effecting the
$\mu$ value and in this sense the high-Z abundances are relatively
uncoupled to the hydrodynamics.  However, because line emission from
these elements, Ne through Fe, can be clearly seen in our X-ray
spectra these abundances are well constrained by the data.  Two other
``degrees of freedom'' in generating the observed X-ray emission from
the hydrodynamics are absorption, both foreground and internal, and
clumping in the ejecta and/or wind~\citep{chugai94}.  Both of these
have little effect on the higher energy flux, 2--8~keV, but can
produce 20\% to 50\% changes in the low-energy flux.  In this low
energy range there is some degeneracy between the CNO abundances, the
$N_H$, the clumping parameters, and the Fe abundance through Fe-L
lines.  For our multi-epoch model we have set the $N_H$ at the HETG-00
and HETG-09 epochs to 9.6 $\times 10^{21}$~cm$^{-2}$ and 7.8 $\times
10^{21}$~cm$^{-2}$, respectively, up to 6$\times 10^{21}$ ~cm$^{-2}$
of which is likely from absorption in the Milky Way and Circinus
Galaxy disks. We then interpolate to other epochs based on the amount
of unshocked CSM.  For the swept-up shell clumping, which likely
extends over a range of values, we have chosen a density factor of 7.0
and a filling fraction of 0.0020; these are sufficient to generate the
Si He-like lines seen at early times.

The output of our calculations provides the emitted X-ray flux versus
time in the 0.5-2 keV band and the 2-8 keV band that we have used, as
well as spectra and spectral model parameter files at each epoch.
Furthermore, we have performed a detailed comparison of the
line-shapes between the data and model Si and Fe lines at the HETG-09
epoch, which are described in the companion paper by Bauer et
al.~(2010).

\begin{figure*}
\includegraphics[angle=0,scale=1.]{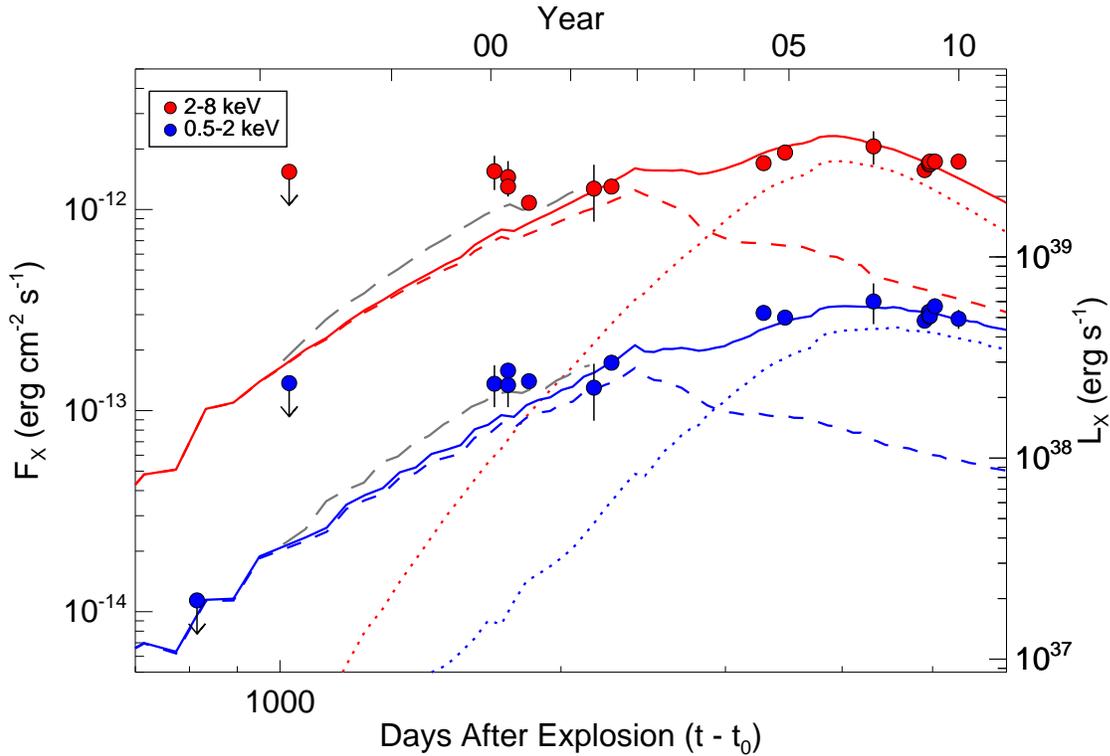}
\caption{The X-ray flux in the soft (0.5-2 keV; blue lines) and hard
  (2-8 keV; red lines) bands computed from our simulations, and
  compared to the observations (circles - error bars are shown).  The
  data are as observed, not corrected for $N_H$ absorption. The
  emission from the higher density ``clumps'' is included, although it
  does not make a significant contribution to the light curve,
  especially at late times.  An explosion date of 1995.4 is used. The
  individual contributions of the shocked ejecta (dotted lines) and
  shocked CSM (colored dashed lines) to the hard and soft fluxes is
  also shown. The CSM dominates the early evolution up to about 7
  years, then the shocked ejecta begin to take over and dominate the
  later evolution. The abundances used for the figure are given in
  Table 1. The long-dashed grey lines indicate the light curve that
  would be obtained if the density profile depicted by the thin solid
  line in Figure \ref{fig:hydroinit} was used, showing how a slight
  rearrangement of the mass in the shell can affect the early-time
  light curve.
\label{fig:xrayflux}}
\end{figure*}

Figure ~\ref{fig:xrayflux} shows the comparison between the flux
calculated from our hydrodynamic simulations and the observations. The
abundance list used to generate these plots is given in Table 1.  The
separate contribution of the shocked ejecta and the shocked CSM to the
flux are also shown. The modelled flux is within 20\% except for the
year 2000 data where it is low by 30-50 \%.  As Figure
\ref{fig:hydroevol} (panel 3, 5.02 years) shows, at this time the flux
is determined by the inner-half of the dense shell. We have verified
(Figure \ref{fig:xrayflux}) that shifting ~10\% of the dense shell
mass from the outer half to the inner half can produce the needed flux
increase with little change at later times.  While it is tempting to
expand the shape of the dense shell into a set of basis functions and
adjust them for the best light curve we feel this would be ``over
tuning" our simple model and limited data.  Certainly in future when
IXO will obtain a deep, well-sampled dataset of a similar SN, such an
approach could be very fruitful.

It is clear that for about the first 7 years the shocked CSM material
dominates the X-ray lines and the continuum, but after that the
contribution of the ejecta increases as the forward shock exits the
shell. By about 8 years the shocked ejecta becomes the primary
contributor. If our approximation of a wind medium exterior to the
dense shell is correct, then the shocked ejecta will remain the
primary contributor to the total flux for more than 25 years. It is
interesting to note that this is exactly the opposite scenario as
deduced for another SN with a high X-ray flux, SN 1993J
\citep{nymarketal2008, chandraetal2009}, where the reverse-shocked
ejecta was found to dominate the X-ray emission for the first several
decades. The dominant early contribution of the CSM is clearly due to
the presence of the bubble and dense shell in this case.

Our method enables us to compute simulated spectra using
non-equilibrium ionization, which we can compare to the
observations. Observed {\it Chandra} HETG spectra are available for
2000, 2004, and 2009, and observed XMM spectra for 2001.  One
important point of our method is that {\em the same hydrodynamical
  simulation can produce spectra that match the observed ones at 4
  different epochs}, without any unnecessary free parameters.  Given
the transition from CSM to ejecta dominance in the emission, the CSM
higher-Z abundances were obtained by using fits to both the HETG 2000
and XMM 2001 spectra, while the ejecta abundances were set by fitting
to our 2009 spectrum.

\begin{figure*}
\includegraphics[angle=0,scale=0.7]{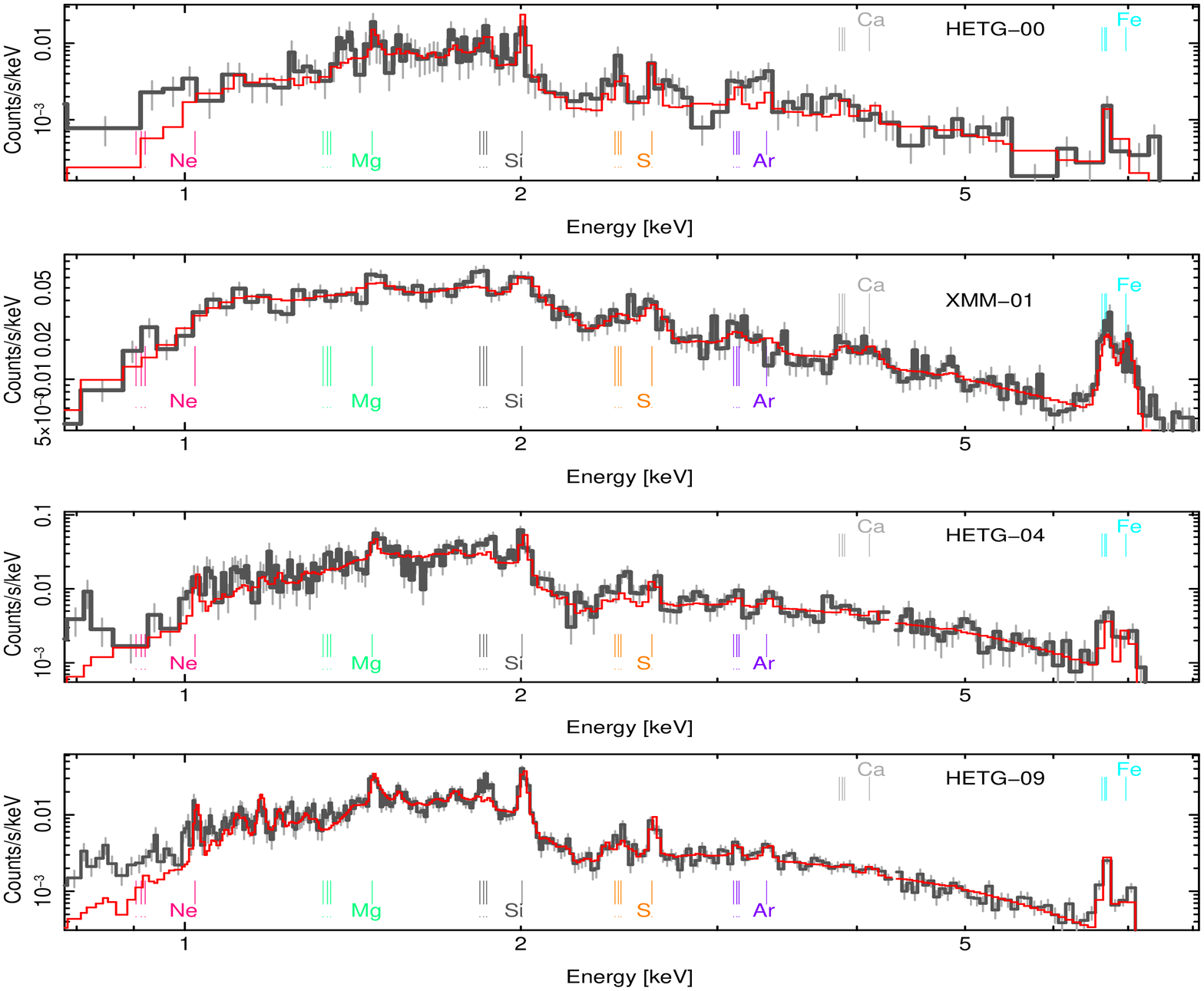}
\caption{Comparison of the observed (black) and calculated (red)
  spectra at various epochs. H and He-like transitions of various
  elements are marked on the plot. Our simulated spectra are able to
  match the continuum beautifully, and the comparison with various
  lines is adequate. A single normalization factor is adjusted at each
  epoch - the spectra do not contain any other free
  parameters. Further details of how the spectra were calculated are
  given in \S 4.
\label{fig:96crspec}}
\end{figure*}

Figure~\ref{fig:96crspec} shows the comparison of the simulated and
observed spectra at all the 4 epochs at which observed spectra are
available. Our simulated spectra are able to fit the continuum
emission very well, as well as most of the major lines (H and He-like
transitions of the various lines are shown on the plot to aid the
reader). We also manage to fit the line transitions in the hard X-ray
band adequately. Initially, our simulated spectra did not adequately
reproduce the He-like transitions of many elements, especially those
that occur in the soft X-ray band below 2 keV. In order to fit these
better, we assumed that there existed a component, about 1\% by mass,
with a density that was about an order of magnitude higher than that
of the surrounding medium, and whose temperature was consequently an
order of magnitude lower in order to preserve pressure
equilibrium. High-density clumps may be responsible for this
component. The rich and complicated line structure, at both the X-ray
and optical wavelengths, is a good indicator of clumping, and the
small percentage required to fit the various lines is well within the
realm of possibility.

\section{SN1996cr as a Type IIn - Clumps and narrow-line formation} SN 1996cr 
was classified as a Type IIn supernova, based on the presence of
narrow lines in the VLT spectrum from 2006 \citep{baueretal2008}. In
the model presented herein, by 2006 the shock had exited the shell and
was expanding in the external wind region. Thus the narrow optical
lines could not arise due to the shock within the shell, whose maximum
velocity in any case was much higher than the $\sim 670\, {\rm km}\,
{\rm s}^{-1}$ FWHM of the narrow component \citep{baueretal2008}. It
is possible that the actual FWHM could be even smaller and the lines
even narrower, beyond the limit of resolution of that particular
observation.

Could the narrow lines arise from recombination in the unshocked,
ionized CSM, which we suggest (below) is a RSG wind? The velocity of
the putative RSG wind would be low, $\la 50\, {\rm km}\, {\rm
  s}^{-1}$, and this line would not be resolved. In order to get the
observed flux in the narrow lines, the RSG wind must have a certain
minimum mass. Using equation 11 in \citet{Salamancaetal98}, we find
that in order to get the required H$\beta$ luminosity, a H number
density in excess of 10$^8 \,{\rm cm}^{-3}$ is required at the
position of the shock, at the time the VLT spectrum was taken (Jan
2006). This is more than 3 orders of magnitude higher than the density
in our simulations, suggesting that in our scenario this line could
not have come from the unshocked CSM.

It is possible that the lines arise from shocked dense clumps
suggested in \S \ref{sec:datacomp}. In order to see if this
explanation is feasible, and to estimate the density, number and
filling fraction of clumps, we follow the method outlined by
\citet{chugai94}. From the simulations, the average pressure in the
shocked region at 10.5 years is about 3.9 $\times 10^{-3}$ dynes
cm$^{-2}$. If the shock transmitted into the clump has a maximum
velocity $v_c$ of 670 km s$^{-1}$ \citep[the approximate velocity of
  the narrow H$\alpha$ line;][]{baueretal2008} and is in pressure
equilibrium, then the clump density ${\rho}_c$ is 8.69 $\times
10^{-19}$ g cm$^{-3}$. Using the average value of the mean molecular
weight for the CSM (see Table 1), this would mean a number density
$n_c \sim 6.9 \times 10^5 {\rm cm}^{-3}$. This density is consistent
with the ``clumps'' assumed in the previous section.

The cooling time of the shock in the clumps must be small compared to
the destruction time for the clumps by Rayleigh-Taylor and
Kelvin-Helmholtz instabilities. The temperature of the shock driven
into the clumps, following \citet{chugai94}, is $T_c = 1.36 \times
10^7\,v_{c_8}^2 \sim 6.1 \times 10^6 $ K. The cooling function at that
temperature, from \citet{chevalier1994}, has a value of 5.26 $\times
10^{-23}$ ergs cm$^3$ s$^{-1}$. Assuming that the cooling time
is less than the time-scale on which the instabilities develop ($\sim
\, a_c/v_c$) gives a minimum size for the clumps of $a_c \sim 1.55
\times 10^{15}$ cm.

As suggested by \citet{chugai94}, the narrow component arises in the
cool material behind the clump shock, which radiates due to the
reprocessing of X-ray emission from the hot shocked gas. As the shock
cools, it will progressively radiate X-ray, then UV and then optical
emission from the cooling region behind the cloud shock. An upper
limit to the luminosity of the ensemble of $N_c$ shocked clumps is
given by:

\be
L_{H\alpha} = 0.25\, \pi \, a_c^2\, {\rho}_c\,v_c^3\, N_c
\ee

The luminosity of the narrow component of H$\alpha$ is 2.79 $\times
10^{38}$ ergs s$^{-1}$. Using the values of ${\rho}_c$ and $v_c$ above
gives

\be
 N_c \sim 1359\,{a_c}_{15}^{-2}
\ee

\noi where ${a_c}_{15}$ is the clump size in units of 10$^{15}$
cm. Keeping in mind the lower limit on $a_c$, we adopt $a_c \sim
10^{16}$ cm, which gives $N_c \sim 13.6$. The total mass in clumps is
about 0.025 $\msun$. Note that the (3D) filling fraction of the
clumps, $f_c \sim 1.2 \times 10^{-3}$, is quite low, and consistent
with that obtained from the X-rays given the large uncertainties. The
(2D) covering factor $\sim$ 0.027, thus indicating that they will not
obscure the SN. However these clumps could add additional column
density along certain sightlines.

A consistency check on the above calculations can be done by comparing
the ratio of the velocities. Since the clumps are in pressure
equilibrium with the interclump medium, the ratio of the interclump to
clump shock velocity must be proportional to the square-root of the
clump to interclump density. The density of the unshocked gas at this
epoch in the simulation is 1.34 $\times 10^{-20}$ g cm$^{-3}$. Given
the density of the clumps mentioned above, the jump in density between
the clumped and interclump medium is a factor of 65. If these are in
pressure equilibrium, then the velocity ratios must be equal to the
square-root of this, or about 8.06. Therefore, if the velocity of the
shock within the clumps is 670 km s$^{-1}$, this implies that the
shock velocity must be about 5400 km s$^{-1}$. This is close to the
velocity of the CSM shock in our simulations (see figure
\ref{fig:shkvel}), as would be expected. However, this is not the
velocity of the broad H$\alpha$ line mentioned in
\citet{baueretal2008}, which would be assumed to reflect the shock
velocity. The inference is that either the broad H$\alpha$ velocity
has been underestimated, or that it does not arise from the shocked
CSM, but perhaps from the shocked ejecta. Computing the FWHM of
H$\alpha$ depends on fitting the continuum precisely, which depends on
the degree and nature of the reddening towards the SN, which is not
accurately known. It is possible that the H$\alpha$ has been
underestimated, and that a more accurate fit to the H$\alpha$ line
will yield a FWHM compatible with this value. 

The computed values for the number, size and density of clumps are
comparable to those found by \citet{chugai94}, and are consistent with
the ``clumps'' referred to in the previous section. The notion that
the narrow optical lines arise in dense shocked CSM clumps, which is
the same population that is also responsible for the lower-temperature
(clumped) X-ray lines, is quite appealing. However the H$\alpha$ lines
are formed at the systemic velocity, while the X-ray lines are
generally shifted by 3000-5000 km s$^{-1}$ from the systemic velocity.
Therefore, it appears that two separate populations may be
needed. This will be investigated further in the companion paper by
Bauer et al.~(2010).

\section{Discussion}
\label{sec:discussion}

\subsection{Implications for the SN progenitor}
\label{sec:prog}
What does our model suggest for the hydrodynamic evolution of the
progenitor star? The derived density profile, with a low-density
interior surrounded by a dense shell, and a wind region exterior to
it, resembles a wind-blown bubble \citep{Weaver1977} formed by the
interaction of two winds. The formation of wind-bubbles, and the
evolution of a SN within such wind-bubbles, has been extensively
studied in the past \citep{cd89, Chevalier1989, tenorioetal1990,
  tenorioetal1991, francoetal1991, Dwarkadas2005, dwarkadas2007a,
  Dwarkadas2007d}. These calculations have suggested that the
interaction of the SN shock wave with the dense shell delineating the
boundary of the wind bubble leads to increasing X-ray and radio
emission. This is evidently the scenario that best resembles the
emission from SN 1996cr. This is only the second case after SN 1987A
where a sustained increase in the X-ray and radio light curves over a
multi-year period is seen. However, the shell in the case of SN 1996cr
is 6 times closer to the star than the dense equatorial ring
surrounding SN 1987A. Its density in our model is a few times higher
than the ring density deduced for SN 1987A. The most recent data
suggest that the shock has exited the shell and the emission is
turning over. Thus SN 1996cr provides an early glimpse into what the
further evolution of the lightcurves in SN 1987A will look like, as
the SN shock in that case is just beginning to interact with the dense
equatorial ring of material \citep{racusinetal2009}. In this picture,
SN 1996cr represents a highly compressed, and much brighter version of
SN 1987A.

If we assume that the dense shell consists only of swept-up wind
material, then the mass of the shell is equal to the mass of swept-up
wind material. We can apply the wind bubble scenario to this model. In
this scenario a fast supersonic wind sweeps up the wind from an
earlier phase into a thin, dense shell. We denote the earlier phase by
subscript 1 and the later phase by subscript 2. The mass of the
swept-up shell is about 0.64 $\msun$. If this is made up of swept-up
wind material, then M$_{shell} = ({\mdot}_1 / v_1) R_{shell}$. Given
that the shell outer radius is 1.5 $\times 10^{17}$ cm, this gives for
the external wind a value of $B_1 = {\mdot}_1 / v_1$ = 8.5e15 g
cm$^{-1}$. If we assume that the wind velocity $v_1$ is 10 km
s$^{-1}$, as is appropriate for a red supergiant wind, then the
mass-loss rate of the star in this phase would be ${\mdot}_1 \sim 1.33
\times 10^{-4} \msun \,{\rm yr}^{-1}$, which is again reasonable for a
RSG wind. Thus one possibility is that the star in this phase may have
been a RSG.  If we assume a larger wind velocity, then the mass-loss
rate must scale accordingly. Mass-loss rates on the order of 10$^{-3}
\msun\,{\rm yr}^{-1}$ have been deduced for hypergiant stars with a
wind velocity of about 40 km s$^{-1}$ \citep{humphreysetal97}, and it
is possible that the star in this phase could have been a hypergiant,
consistent with the possibility that the progenitor was a very massive
star. Even higher velocities would make it unreasonably large for any
phase except perhaps the outburst stage of a Luminous Blue Variable
(LBV) phase \citep{hd94}. But such large mass-loss rates are
unsustainable for a large period of time, and we would expect such a
phase to be necessarily short-lived \citep{hd94}. S Doradus type
instabilities generally have a timescale of years to decades
\citep{avg01}. Therefore the existence of an LBV phase would be
apparent if the density profile were to change rapidly as the shock
expanded outwards, which would show up as a strong deviation in the
X-ray light curve from that suggested by us in this paper. We look
forward to such measurements in future.

\begin{figure*}
\includegraphics[angle=270,scale=0.75]{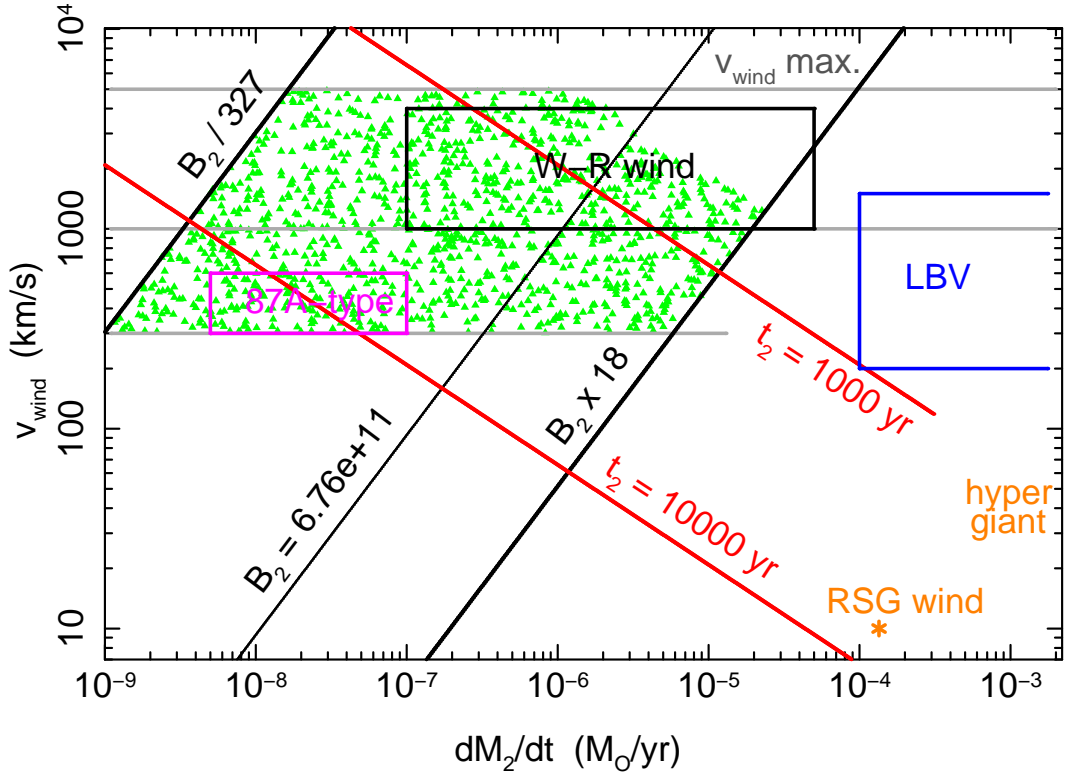}
\caption{The range of mass-loss parameters for the pre-SN wind,
  assuming the wind density parameters B$_1$ (denoted by an orange
  star symbol) and B$_2$ (see \S \ref{sec:prog}), and timescales
  between 1000 and 10$^4$ years.  Possible values lie at the
  intersection of the B$_2$ and $t_2$ lines. As can be seen, the range
  of theoretically possible values that give realistic wind parameters
  is limited, and is shown populated by the green triangles. Various
  portions of the diagram that correspond to mass-loss parameters for
  W-R stars, SN 1987A-type BSGs, LBVs, RSG and Hypergiant stars are
  marked. The most likely progenitor is a W-R star.
\label{fig:windparam}}
\end{figure*}

What about the final phase of the wind before the star exploded?  Only
upper limits to the X-ray emission are available for this period, thus
only providing an upper limit to the density profile. In the
simulations, we assume a wind with a value of $B_2 = \mdot / v_w$ =
6.76 $\times 10^{11}$ g cm$^{-1}$, which we refer to as the standard
value of $B_2$. This means that for a wind velocity of a 1000 km
s$^{-1}$, the mass-loss rate is on the order of 10$^{-6} \msun {\rm
  yr}^{-1}$ (Figure \ref{fig:windparam}).  The wind density is most
constrained by the upper limits on the X-ray emission between days
600-900. A factor of 18 increase in the value of B$_2$ delays the
high-flux turn-on and increases the discrepancy for the year 2000
measurements by 20 to 30 \% . Thus the maximum mass-loss rate that
would be able to produce a reasonable light curve (within 30\% of the
observed values) and just about satisfy other constraints (for a
velocity of 1000 km s$^{-1}$) would be about 1.8 $\times 10^{-5} \msun
{\rm yr}^{-1}$.

At the lower end we find that the density could be significantly
smaller before the shock collides with the dense shell too early and
exceeds the upper limits on days 700-900 (although this can be
somewhat mitigated by changing the explosion date, which is uncertain
by about a year). For the same 1000 km s$^{-1}$ velocity, we find a
lower-limit to the mass-loss rate of about 3.26 $\times 10^{-9} \msun
{\rm yr}^{-1}$. Thus, although we cannot exactly constrain the inner
density, we find that the best fit light curve leads to mass-loss
rates between 3. $\times 10^{-9} \msun {\rm yr}^{-1}$ and 2. $\times
10^{-5} \msun {\rm yr}^{-1}$, for wind velocities of 1000 km s$^{-1}$.

At the higher end, this mass-loss rate would be compatible with a
Wolf-Rayet star progenitor for the SN. This would make the SN a Type
Ib/c.  At the lower end, and especially for lower velocities, the
mass-loss rate and wind velocity combination approaches that deduced
for SN 1987A \citep{chevalier1995, lundqvist1999, Dwarkadas2007b,
  Dwarkadas2007d}, and may further strengthen the analogy between SN
1996cr and SN 1987A. While we are not suggesting that the progenitor
of SN 1996cr was a blue supergiant, it lies within the realm of
possibility.

One implication of these low mass-loss rates is that it is unlikely
that the progenitor star was an LBV star. LBVs are highly luminous,
unstable stars, which can undergo episodes of dramatic mass-loss
\citep{hd94}, leading to the formation of circumstellar nebulae around
the star. LBV nebulae are generally of order 1pc \citep{weis01}, much
larger than the circumstellar nebula in our simulations. LBV stars
have often been suggested as progenitors for Type IIn SNe
\citep{galyam2007, smith08, vink2008, galyam2009, milleretal2009},
based mainly on the fact that they undergo mass-loss episodes with a
high mass-loss rate $\ga 10^{-4} \msun {\rm yr}^{-1}$. LBV winds are
also slower than W-R winds, with velocities on the order of 200 km
s$^{-1}$ \citep{vk07}. Assuming such a velocity would lower the
mass-loss rates derived above by a further factor of 5. This would
lead to mass-loss rates lying between about 6 $\times 10^{-10} \msun
{\rm yr}^{-1}$ and 4 $\times 10^{-6} \msun {\rm yr}^{-1}$ (see Figure
7) for a presumed LBV progenitor velocity. These are significantly
lower than mass-loss rates proposed for LBV progenitor stars.
Mass-loss from an LBV in the quiescent stage may be consistent with
the higher end of the rates deduced here \citep{hd94}, but then it is
impossible to distinguish this from other massive supergiant stars
\citep{Stothers1996} based purely on hydrodynamic considerations. In
any case by assumption this would indicate that the wind bubble was
not the result of an LBV outburst.

We can further constrain the wind parameters by including the shell
formation time, t$_2$. As above, we assume that the shell was formed
by the interaction of two winds with constant mass-loss parameters,
the first wind being the earlier RSG wind, followed by the second wind
phase. The shell radius is given by R$_{shell} = (L_2 /[1.5
  B_1])^{1/3} t_2$ cm \citep{dwarkadas1997} where $L_2 = 0.5 {\mdot}_2
{{v_w}_2}^2$ is the mechanical luminosity of the second wind. With
R$_{shell}$ = 1.5 $\times 10^{17}$ cm, and $B_1 = 8.5 \times 10^{15}$
g cm$^{-1}$, we still have one more unknown, the age of the shell. For
an age $t_2$ of 10,000 years, we obtain $L_2$ = 1.37 $\times 10^{33}$
ergs s$^{-1}$. This is on the weak side for a W-R star, but may be
comparable to that for the BSG progenitor of SN 1987A. Using the
standard value of $B_2$ above, this gives a wind velocity ${v_w}_2$ of
160 km s$^{-1}$, and mass-loss rate of the second wind about 1.7
$\times 10^{-7} \msun {\rm yr}^{-1}$.  For the lowest value of $B_2$
we get a velocity of 1098 km s$^{-1}$ and mass-loss rate of 3.6
$\times 10^{-9} \msun {\rm yr}^{-1}$, while for the highest value we
get a velocity of 61 km s$^{-1}$ and a mass-loss rate of 1.17 $\times
10^{-6} \msun {\rm yr}^{-1}$.  Values of $B_2$ slightly lower than the
standard value are comparable to the BSG progenitor of SN 1987A, while
other combinations do not seem reasonable for known stars. Thus an age
of about 10$^4$ years is possible only for low values of $B_2$,
leading to a low velocity and low mass-loss rate progenitor. The
expansion velocity of the shell R$_{shell} / t_2$ is about 5 km
s$^{-1}$ (which technically requires that the RSG wind velocity be
also reduced, but factors of a few are easily accommodated).

It can be seen that $L_2 \propto t^{-3}$, and therefore if we reduce
the age t$_2$ to 1000 years, we get $L_2$ = 1.37 $\times 10^{36}$ ergs
s$^{-1}$. This implies a mass-loss rate ${\mdot}_2$ of 1.7 $\times
10^{-6} \msun {\rm yr}^{-1}$ and a wind velocity ${v_w}_2$ of 1594 km
s$^{-1}$ for the standard value. The expansion velocity of the shell
now is around 48 km s$^{-1}$. These parameters are reasonable for a
W-R star. It can be seen that lowering the values of $B_2$ by more
than a factor of 30 increases the wind velocity to $>$ 5000 km
s$^{-1}$, which is incompatible with known wind velocities.  The
highest value of $B_2$ gives a velocity of 608 km s$^{-1}$ for a
mass-loss rate of 1.16 $\times 10^{-5} \msun {\rm yr}^{-1}$.  In this
case it seems that values ranging from about 3\% of the standard value
to 4 times the standard value give mass-loss parameters compatible
with known stars.

If we reduce the age $t$ still further, thereby increasing $L_2$, the
wind velocity increases proportionally.  Reduction in age by a factor
of 2 gives a mass-loss rate of 3.4 $\times 10^{-6} \msun {\rm
  yr}^{-1}$ for a wind velocity of 3188 km s$^{-1}$ for the standard
value of $B_2$. Values much lower than the standard value are pretty
much ruled out because they give too high wind velocities. At the
higher end this gives a wind velocity of 1216 km s$^{-1}$ with a
mass-loss rate of 2.33 $\times 10^{-5} \msun {\rm yr}^{-1}$, probably
reasonable for a W-R star.  It is unlikely therefore that the age can
be reduced by even a factor of 2 for the lowest values of $B_2$. For
high values of $B_2$ the age could possibly be reduced by a factor of
3-5.  An age of 250 years with the highest possible value of $B_2$
gives a wind velocity of 2431 km s$^{-1}$ with a mass-loss rate of
4.67 $ 10^{-5} \msun {\rm yr}^{-1}$.  These may just about be
compatible with W-R stars. However we note that ages less than 1000
years gives velocities for the swept-up shell that are much larger
than observed shell velocities around massive stars. Observed shell
velocities of nebulae around W-R stars \citep{cappaetal03}, as well as
around BSG stars \citep{ch91}, are all in the range of 10 km s$^{-1}$
or less. Although much larger shell velocities are seen in LBV nebulae
\citep{weis01}, the mass-loss rate and velocity combinations that are
possible here never resemble those seen in LBV outbursts.  The
conclusion is that the age of the shell lies between 10$^3$ and 10$^4$
years, with a preference for a smaller age.  There is a large range
over which a W-R progenitor seems viable, and a smaller range where a
SN 1987A-like BSG progenitor may work. The range of values for the
mass-loss and wind velocity, given the parameters used and the
acceptable age of the wind-blown shell, are shown in Figure
\ref{fig:windparam}.

The above analysis then suggests the following scenario for the
progenitor: It was a massive O star that evolved off the main sequence
(MS) to become a RSG or hypergiant with a mass loss rate of $\sim
10^{-4} \msun {\rm yr}^{-1}$. This star then entered a second, brief
post-MS phase wherein it became either a BSG star, or more likely a
W-R star.  This second post-MS phase lasted between 10$^3$ and 10$^4$
years, with a preference towards the smaller age for a W-R progenitor,
and a larger age for the 87A-like BSG progenitor.  The BSG case
suggests a star with a zero age main-sequence (ZAMS) of 15-30 $\msun$
\citep{wpe88}.  The W-R case points to a massive ZAMS star $> 30
\msun$, which went from a RSG to a brief W-R phase before
exploding. The abundances derived from the X-ray fitting (\S
\ref{sec:abundances}) suggest a preference for the W-R case.

A final phase lasting for upto 10$^4$ years is not unlike the case of
SN 1987A, where the final BSG phase is assumed to have lasted for
about 20,000 years before the star exploded \citep{mccray2003,
  mccray2007}. The small timescale derived for the final pre-SN phase
puts SN 1996cr in the category of several other recent SNe, including
SN 1987A, SN 2006jc \citep{foley2007, tominagaetal2008}, SN 2006gy
\citep{smithmccray2007}, SN 2008iy \citep{milleretal2009} and SN 1994W
\citep{chugai2004}, which lost about a solar mass of material a short
while before the stellar explosion.  In either case the current
observations and our simulations are helping to shed light on the last
$< 10^4$ years , or $\la 1\%$ of the star's multi-million year
history.

Is it possible that the wind bubble scenario does not apply at all,
that the shell is expanding ballistically, and that it was ejected in
an LBV eruption? Given the speeds of the winds seen in LBV nebulae
\citep{weis01}, and the fact that the shell is expanding
ballistically, one would expect it to have speeds of at least 500 km
s$^{-1}$. The above analysis shows that such an eruption would have
occurred less than a 100 years before the SN explosion. However, if
the shell is purely wind driven, it is not clear why the interior is
such low density, and why the shell is deduced to be so thick. Density
contours from radiatively driven wind models of LBV nebulae can be
seen in Figures 3 and 4 of \citet{db02}, and do not resemble the
density structure derived herein, although it is possible that other
mass-loss mechanisms may result in different density distribution. In
particular the thickness of the shell (50\% of the radius) implies
some kind of wind interaction, and would be hard to explain in a model
where the wind was expanding ballistically. Furthermore, the fact that
the mass in the shell is found to be equal to the mass in the external
wind of the assumed density upto the outer radius of the shell, would
be an extreme coincidence if the shell was not composed of swept-up
external wind material. Therefore we consider this to be a less likely
possibility.

\subsection{Abundances} 
\label{sec:abundances}
\begin{figure*}
\includegraphics[angle=-90,scale=0.65]{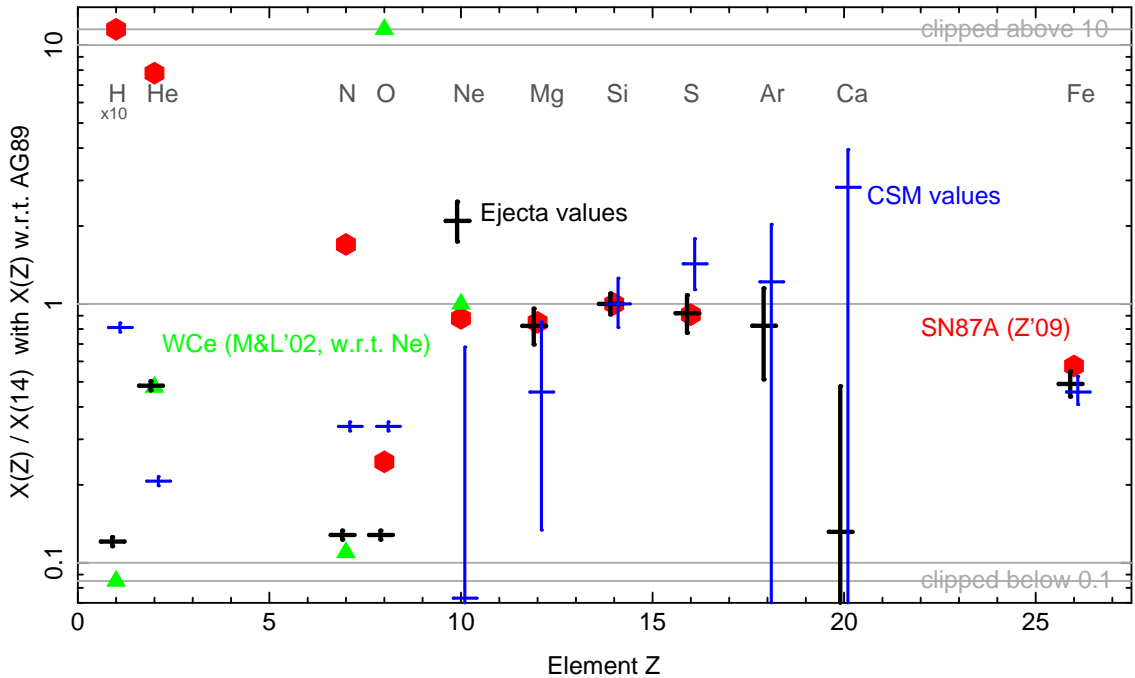}
\caption{Abundances derived from our X-ray fitting, compared to others
  found in the literature. The H, He, N, O values are shown with very
  small error bars to indicate that they are fixed (H, He) or
  undetermined (N,O). The Red hexagons are SN 1987A measurements from
  \citet{zhekovetal09}.  SN 1996cr appears to have 10 times more Z $>$
  8 metals compared to the He (and H) there, and seems to posses a
  H,He-poor or metal-rich CSM compared to SN 1987A.  But the ratio of
  the heavy (Z $\ga$ 8) metals is about the same - the Oxygen value in
  SN 1996cr for instance could be easily set equal to that seen in SN
  1987A without affecting the light-curve fits.
\label{fig:abund}}
\end{figure*}

Table 1 describes the abundances that we have derived in order to
compute spectra that match the observations. The statistical
1-$\sigma$ range of values for each abundance is also given.  Figure
\ref{fig:abund} graphically displays the range of abundances deduced
from our X-ray fitting. The H, He, N, O values are shown with very
small error bars to indicate that they are fixed (H, He) or
undetermined (N,O). In the figure we also compare the abundances to
those derived from LETG and HETG observations of SN 1987A
\citep{zhekovetal09}, as well as those given for WCE stars by
\citet{ml02}. These comparisons lead to the following observations:
(1) The X-ray data provide very little constraint on the ratios of He
to H, or $\mu$, of the plasmas.  Agreement with the light curve and
spectra was improved with our choice of a H-rich CSM and a more
He-rich ejecta. However it is likely that other possibilities, e.g.,
both plasmas with a $\mu \sim$ 1.0, could be accommodated with some
adjustments to the density profile. (2) Our values for the relative
abundances of the heavier elements in the CSM are consistent with
those for the CSM around SN 1987A derived from fitting the LETG and
HETG spectra \citep{zhekovetal09}. Those abundances were found to be
consistent with those of the LMC, and with SNRs in the LMC. In the
case of the Circinus galaxy, \citet{olivaetal99} arrive at an average
metallicity of -0.7 $\pm$ 0.3 from considerations of the equivalent
width of the stellar CO lines in a square of 100pc on a side. This
indicates a metallicity lower than that of the LMC, which would
enhance the potential for a BSG to explode.  (3) The main difference
between SN 1996cr and SN 1987A abundances is in the H and He that make
up the ``continuum''. Our fits predict H and He abundances that are an
order of magnitude lower than those derived for SN 1987A. (4) The low
H and He in the ejecta are suggestive of a W-R star which has expelled
its H and He envelope, consistent with the analysis from purely
hydrodynamical considerations above.  The abundance measurements are
closest to those for a WCe star as given in \citet{ml02} (5) The
  CSM in this case, although mainly composed of H and He, is
  overabundant in heavy elements like S and Si. The existence of
  strong Si and S H-like lines, and the very strong Fe H-like line at
  $\sim$ 7 KeV in the 2001 XMM spectrum, requires a metal-rich
  plasma. Although RSGs are not expected to have such metal-rich
  winds, it is interesting to note that the RSG WOH G64 shows the
  presence of [SII] 6717/6731 in its spectrum \citep{lmpo09}, while
  the M supergiant VY CMa shows the presence of Si I lines, and SiO
  maser emission \citep{wg01}, along with various signatures of other
  metal species. Therefore one possibility is that the winds are
  enhanced in Si and S. An alternate possibility is that the shocked
  CSM could actually be a mixture of CSM and ejecta material, due to
  mixing at the interface between the shocked CSM and shocked ejecta,
  which can lead to Rayleigh-Taylor ``fingers'' of ejecta propagating
  into the CSM. In order to test this possibility, we have assumed a
  modified CSM composition, that is made up of equal parts (by number)
  of the SN 1987A CSM \citep{zhekovetal09} and SN 1996cr ejecta. Such
  a 50-50 mixture has a resulting abundance similar to that of our
  ``CSM'', and, if used in place of our CSM abundance values, gives a
  similar light curve and spectra. Thus what we are labeling as
  ``CSM'' could be contaminated by ejecta mixing, and the derived
  abundance reflects this mixture.  (6) The H abundance is somewhat
higher in the CSM while the He abundance is higher in the ejecta. This
may indicate that the star gave off most of its H layer but retained
somewhat more of its He layer, pointing perhaps to a Type Ib SN. The
characteristic velocity computed in \S \ref{sec:hydro} is also closest
to that of a Type Ib SN \citep{maureretal09}. The overall abundance of
metals is still high in the CSM, suggesting perhaps that most of the H
and He was lost in a much earlier stage, and is much further out in
radius.  This is consistent with the fact that the circumstellar
bubble and dense shell with which the SN shock wave was interacting in
its first decade was emitted in the very last stages of the stellar
lifetime, and that much more CSM material that was released earlier is
to be found further out at a larger radius.  If so the SN should probe
this material over the next few decades. (7) One aspect that we
  have not considered, and which may play a role in this scenario, is
  grain chemistry. RSG stars are known to have dusty winds, and stars
  with higher mass-loss rates are more likely to show the presence of
  dust \citep{jones01}. It is possible that the dense shell, if
  composed of RSG wind material, has a significant dust component.
  Under equilibrium conditions, metals can condense in dust grains,
  thus depleting certain species and affecting the X-ray plasma
  abundances.  Alternatively, given the presence of the high density
  shell and the hot gas, which can provide ions with an energy of
  several keV, it is also possible that sputtering of the dust by hot
  gas as the shock travels through the dense shell is important. This
  may alter the X-ray plasma composition.

\begin{table*}
 \centering
 \begin{minipage}{100mm}
  \caption{The ejecta and CSM abundance distributions as deduced from
    spectral fitting.  The abundances are expressed relative to Si
    with respect to the values listed in \citet{ag89}; the 1-$\sigma$
    range for elements with detectable lines are also given.  The
    ``continuum'' abundances, H through O, are not well constrained by
    the data (see text).  Other columns give these abundances
    converted to number fraction and to percent by mass.  For the
    ejecta (CSM) the mean ion mass, $\mu_A$, is 3.548 (1.693) amu and
    the mean mass per particle, $\mu$, is 1.236 (0.755) amu.}
\begin{tabular}{lcccccc}
  \hline
     & \multicolumn{2}{c}{Abund.s w.r.t.\ Si (AG89)}         &
     \multicolumn{2}{c}{Number fraction} &  \multicolumn{2}{c}{Percent mass}\\
Element &   Ejecta       & CSM                       &   Ejecta       & CSM   &   Ejecta       & CSM\\
\hline
 H   & 0.0121            & 0.0811                    & 2.02e-01 &  7.96e-01 & 5.69  &  47.0 \\
 He  & 0.483             & 0.207                     & 7.89e-01 &  1.98e-01 & 90.0  &  46.8 \\
 C   & 0.128             & 0.336                     & 7.77e-04 &  1.20e-03 & 0.263 &  0.849 \\
 N   & 0.128             & 0.336                     & 2.40e-04 &  3.70e-04 & 0.095 &  0.306 \\
 O   & 0.128             & 0.336                     & 1.82e-03 &  2.81e-03 & 0.821 &  2.655 \\
   \hline
 Ne  & 2.09 [1.74\,--\,2.48] & 0.07 [0\,--\,0.68]    & 4.30e-03 &  8.81e-05 & 2.427 &  0.104 \\
 Mg  & 0.82 [0.70\,--\,0.96] & 0.46 [0.13\,--\,0.85] & 5.23e-04 &  1.70e-04 & 0.354 &  0.242 \\
 Si  & 1.00 [0.91\,--\,1.10] & 1.00 [0.81\,--\,1.26] & 5.93e-04 &  3.48e-04 & 0.468 &  0.576 \\
 S   & 0.92 [0.77\,--\,1.08] & 1.43 [1.13\,--\,1.78] & 2.49e-04 &  2.28e-04 & 0.225 &  0.431 \\
 Ar  & 0.82 [0.51\,--\,1.15] & 1.22 [0\,--\,2.03]    & 5.00e-05 &  4.34e-05 & 0.056 &  0.102 \\
 Ca  & 0.13 [0\,--\,0.48]    & 2.82 [0\,--\,3.94]    & 5.04e-06 &  6.35e-05 & 0.006 &  0.150 \\
 Fe  & 0.49 [0.44\,--\,0.55] & 0.46 [0.41\,--\,0.52] & 3.83e-04 &  2.10e-04 & 0.605 &  0.693 \\
\hline
\end{tabular}
\end{minipage}
\end{table*}

\section{Summary}
\label{sec:summary}

We have computed a hydrodynamic model that is able to explain the X-ray
light curves and spectra of the young supernova 1996cr. Our intent was
not to completely explore the parameter space - indeed, this is
practically impossible - but to evaluate the best possible model
consistent with all of the {\it available} data. Our model
incorporates a dense shell of material that lies about 0.03pc from the
star and extends radially for about 0.015pc, with a density of about 8
$\times 10^4$ amu/cc. The density interior to the shell is quite low,
although it is not well constrained by the observations. The density
external to this is well approximated by a wind profile at
present. The interaction of the SN ejecta with this shell and wind is
able to reproduce the increasing X-ray light curves that have
distinguished this supernova, and are also consistent with the
behavior of the radio light curves.

We have computed the X-ray emission from our simulations, using
non-equilibrium ionization conditions, and taking into account the
inequality between the electron and ion temperatures. Our method is
both quick and accurate\footnote{When referring to the accuracy of the
  flux obtained from spherically symmetric simulations of a SN which
  is still unresolved at all wavelengths, we mean to better than about
  30\%.} and allows us to compute simulated spectra at every epoch,
that can be directly compared with the observations. We have compared
both the full spectrum as well as specific lines. This is better
described in the companion paper (Bauer et al.~2010, in
preparation). The flux at various epochs, and the computed X-ray
spectra, agree well with the observed data, thereby validating our
model.

Our hydrodynamic model suggests that the progenitor of SN 1996cr was
more likely a BSG or W-R star, with a fast, low density wind. The
interaction of this wind with a prior wind, possibly from a RSG phase,
gave rise to the wind-blown shell with which the SN shock wave was
colliding in its first decade, leading to the increasing radio and
X-ray emission. The similarity to SN 1987A in many different aspects
is quite striking. An LBV progenitor is inconsistent with much of the
data.

Given the similarity with the medium around SN 1987A, which has a
clear ring structure surrounding it, it is interesting to ask whether
a ring structure, rather than a fully developed spherical shell, would
be a possible scenario for SN 1996cr. In our study, we have found that
a somewhat denser shell (about 50\% higher density), coupled with the
assumption that the emission was arising from a region of 2$\pi$
steradians, would give approximately the same level of flux. Thus it
is possible that a donut-shaped structure rather than a spherical
shell could be accommodated, although we have not investigated the
detailed spectra. It seems unlikely that a denser shell would fit
because then the VLBI radius constraint would not be
satisfied. Further investigations in this direction are left to a
future paper as they would require far more data for comparison than
is currently available.

It is possible that the further expansion of the shock may not agree
with our supposition that the medium outside the shell is a RSG wind,
or indeed any kind of wind. Only further data on the expansion can
shed light on this, and we eagerly look forward to continually
exploring the X-ray emission of this interesting and enigmatic object
with the {\it Chandra} and {\it XMM-Newton} telescopes e.g. through
yearly flux monitoring and and future HETG spectroscopy if
feasible. Similarly, a second VLBI measurement, which will provide a
second radius measurement and a direct indication of the current shock
velocity, is eagerly awaited. Finally, we also have applied for
further optical and IR observations that will help to create a
multi-wavelength history of this SN, and enable us to understand the
various properties of this fascinating object.

The relationship of SNe to their progenitor stars is quite uncertain,
as mentioned in the introduction. Surveys that try to pin down the
progenitors are doing a fantastic job \citep[see][and references
  within]{smartt2009}, but the problem is difficult. Theoretical
considerations like those used in this paper provide another
completely different technique to explore the CSM around the
supernova, and link it to its progenitor star. No doubt there is
uncertainty in this, as in any other method that is trying to
ascertain the properties of a star that no longer exists. But by using
the same hydrodynamic simulation to fit multi-epoch X-ray spectra as
well as light curves without adding too many free parameters, we have
tried to mitigate uncertainties as far as we can. The fact that the
simulations reproduce the observed shape of the continuum at each
epoch without any additional parameter speaks for the validity of our
model. It is only by amalgamating the results from various methods
that we can hope to understand the true progenitors of the various
types of SNe. In the case of a SN such as 1996cr, where the
possibility of looking for a progenitor star at the same location,
which faded away after the SN explosion, no longer exists, a
theoretical analysis may be the only method of excavating the SN
progenitor.

\section*{Acknowledgments}
VVD would like to thank Poonam Chandra, Nikolai Chuagi and Don York
for suggestions. We are extremely grateful to the referee, Roger
Chevalier, for a thorough reading of the manuscript, and for several
useful comments that have helped to improve this paper. Support for
this work was provided by the National Aeronautics and Space
Administration through Chandra Awards Number SAO GO8-9074X,
GO9-0086A/B/D and GO0-11095A/B issued by the {\it Chandra} X-ray
Observatory Center, which is operated by the Smithsonian Astrophysical
Observatory for and on behalf of the National Aeronautics Space
Administration under contract NAS8-03060. DD was supported by NASA
through SAO contract SV3-73016 to MIT for support of the Chandra X-ray
Center (CXC) and Science Instruments.

\section{Appendix - Calculation of the Progenitor Wind Parameters}

A detailed description of the wind parameters in an interacting winds
scenario, for various shock strengths and other variables, is given in
\citet{ci83}. A simplified description, as outlined below, is
sufficient for our purposes.

We start with the equation for the radius of the shell in \S \ref{sec:prog}

\be
R_{shell} = (L_2 /[1.5 B_1])^{1/3} t_2
\ee

\noi
Noting that $L_2 = 0.5 {\mdot}_2 {v_w}_2^2$ and ${v_w}_2 =
{\mdot}_2/B_2$ we get

\be
R_{shell} = \frac{{\mdot}_2 t_2}{(3 B_2^2 B_1)^{1/3}}
\ee

which gives

\be
B_2 = \sqrt{\frac{{({\mdot}_2 t_2})^3}{R_{sh}^3 3 B_1}}
\ee

Note that ${\mdot}_2 t_2$ is the total mass of the emitted wind over
time $t_2$. This includes the mass of the freely expanding wind
$M_{fw}$, which extends upto the wind termination shock $R_t$ ($M_{fw}
= B_2 * R_t$) and the mass of the shocked wind, which stretches from
$R_t$ to the inner boundary of the shell $R_i$. If we assume that the
density of the shocked wind is constant throughout, a reasonably good
approximation, and that the wind termination shock is strong,
i.e. with a shock jump of 4, then the density of the shocked wind is 4
times that of the free wind at $R_t$, ${\rho}_{sw} = B_2 / (4 \pi
R_t^2) * 4$, and the mass of shocked wind

\be
M_{sw} =  \frac{B_2}{( \pi R_t^2)} * \frac{4 \pi}{3} ({R_{i}^3 - R_t^3}) 
\ee

\noi
Therefore, with ${\mdot}_2 t_2 = M_{fw} + M_{sw}$ we get an equation for $B_2$

\be 
1 - \sqrt{\frac{B_2[ R_t (1 + \frac{4}{3} ({\frac{R_{i}^3}{R_t^3}-1}))]^3}{3\,R_{sh}^3 \, B_1}} = 0
\ee

This equation can be solved using standard techniques for all $R_t$
between 0 and $R_i$ to get appropriate values of $B_2$. 

Our calculations constrain the density of the shocked wind
${\rho}_{sw}$, so we compute the values of $R_t$ and $B_2$ that
provide us with the constrained values of ${\rho}_{sw}$, and thus set
limits on the progenitor wind density.

\bibliographystyle{mn2e}
\bibliography{paper}


\bsp

\label{lastpage}

\end{document}